%% file: main.tex
\documentclass[preprint]{elsarticle}
\usepackage[margin=2.5cm]{geometry}

\usepackage{lineno,hyperref}
\modulolinenumbers[5]

\journal{Elsevier}

\usepackage[english]{babel}
\usepackage{subfig}
\usepackage{xspace}
\usepackage{upgreek}

\def\dV {\ensuremath{\rm \Delta V}\xspace}
\newcommand{\dVeq}[1]{\ensuremath{\rm \Delta V=#1\,V}}
\def\vbd{\ensuremath{\rm{V_{BD}}}\xspace}
\def\vbdI{\ensuremath{\rm{V_{BD}^I}}\xspace}
\def\vbdAmp{\ensuremath{\rm{V_{BD}^{Amp}}}\xspace}
\def\vbdInt{\ensuremath{\rm{V_{BD}^{Int}}}\xspace}
\def\vbdG{\ensuremath{\rm{V_{BD}^G}}\xspace}
\def\vbias{\ensuremath{\rm{V_{bias}}}\xspace}
\def \degC {$^{\circ}\mathrm{C}$\xspace}

\def\pox{$\rm{p_{DiXT}}$\xspace}
\newcommand{\poxeq}[1]{\ensuremath{\rm{p_{DiXT}=#1\,\%}}}

\newcommand{\pdxeq}[1]{\ensuremath{\rm{p_{DeXT}=#1\,\%}}}

\def\trec{$\rm{\tau_{rec}}$\xspace}
\newcommand{\treceq}[1]{\ensuremath{\rm{\tau_{rec}=#1\,ns}}}
\def\tlong{$\rm{\tau_{long}}$\xspace}
\newcommand{\tlongeq}[1]{\ensuremath{\rm{\tau_{long}=#1\,ns}}}
\def\tshort{$\rm{\tau_{short}}$\xspace}

\def\tap{$\rm{\tau_{AP}}$\xspace}

\def\mm2{$\rm{mm^2}$\xspace}
\def\RQ{$\rm{R_Q}$\xspace}
\newcommand{\RQeq}[1]{\ensuremath{\rm{R_Q=#1\,k\Omega}}}
\def\Plum{$\rm{P_{lum}}$\xspace}
\def\fdcr{$\rm{f_{DCR}}$\xspace}

\newcommand{\Neq}[1]{\ensuremath{#1\cdot 10^{11}\,\rm{MeV\,n}_{\rm{eq}}/\rm{cm}^2}}

\begin{document}

\begin{frontmatter}

\title{Characterisation of silicon photomultipliers based on statistical analysis of pulse-shape and time distributions}

\author[mainaddress]{O. Girard}
\author[mainaddress]{G. Haefeli\corref{correspondingauthor}}
\cortext[correspondingauthor]{Corresponding author}
\ead{guido.haefeli@epfl.ch}
\author[mainaddress]{A. Kuonen}
\author[mainaddress]{L. Pescatore}
\author[mainaddress]{O. Schneider}
\author[mainaddress]{M. E. Stramaglia}
\address[mainaddress]{Laboratory for High Energy Physics, Ecole polytechnique f\'ed\'erale de Lausanne (EPFL), BSP - Cubotron, 1015 Lausanne, Switzerland}

\begin{abstract}
A detailed and accurate characterisation of silicon photomultiplier detectors is required for a better understanding of the signal and noise in many applications. The collected information is a valuable feedback to the manufacturers in their attempt
to improve the performances.
In this paper, we provide a detailed description of how to characterise these photo-detectors. The correlated noise probabilities, the important time constants and the photon detection efficiency are obtained with a statistical analysis of pulse-shape and time distributions. The method is tested with different detectors. The quench resistor, the breakdown voltage and the dark count rate are measured from IV characteristics.
\end{abstract}


\end{frontmatter}


\input{introduction}

\input{experimental_methods}
\input{pulse_shape_analysis}

\input{gain_measurement}

\input{IV_measurement}

\input{pde_measurement}

\input{conclusion}
\section*{References}
\bibliography{main}

\end{document}

%% file: introduction.tex
\section{Introduction}
Silicon Photomultipliers (SiPMs) have become increasingly popular in high energy physics and medical applications during the last decade. The accurate characterisation of these photo-detectors can be difficult depending on the conditions and detector type. For detectors with relatively large pixels and therefore high gain, we have developed methods to measure all important parameters based on the analysis of the pulse waveform.
The most important parameters are the photon detection efficiency (PDE), the noise level and its composition.
There are two categories of noise: random noise produced by thermal excitations in the silicon bulk which are at the origin of the dark count rate (\fdcr), also referred to as primary noise, and noise generated in correlation with a primary noise pulse, called correlated noise.

In our measurements, we record the waveforms of the pulses generated by the photo-detector under test read out with a fast preamplifier and an oscilloscope. The device is protected
against electromagnetic interference by a shielded box.
The excellent signal to noise ratio achieved in this way allows to operate and characterise the detector over a large range of operation voltages without being dominated by the electronic noise.

The characterisation of SiPMs needs to be performed at a given constant temperature.
Temperature dependant parameters as the \fdcr or the breakdown voltage (\vbd) can be corrected if the measured temperature deviates from the nominal value.
For devices with high \fdcr due to irradiation or very large surface, the operation temperature can be lowered in order to reduce the probability of random overlapping dark pulses.
Constant temperature operation can be achieved by either stabilising the temperature with a cooling system or by recording all measurements in a short time. The latter strategy is adopted for most of the characterisations described in this paper.

The characterisation is performed by a scan of the bias voltage (\vbias) over an appropriate range, from which \vbd is determined. The results are given as a function of the over-voltage ($\rm{\Delta V} = \rm{V_{bias}} - \rm{V_{BD}}$). For each \dV value the waveform analysis allows to measure the correlated noise probabilities. The measurement of the SiPM PDE requires a correction to account for the correlated noise. The analysis of the waveforms allows to introduce such corrections with an excellent accuracy within a large range of \dV.
In addition, the waveform analysis allows to measure the gain, the long pulse decay time constant and the pixel recovery time.

\subsection{The SiPM for the LHCb SciFi tracker}

Our team has the responsibility for the development of the SiPM multichannel array used in the LHCb tracker~\cite{TDR} based on scintillating fibres (SciFi).
The tracker will use a total of 590k SiPM channels grouped in 128-channel arrays. The final selected array, produced by Hamamatsu, is referred to as \textit{H2017} in this document.
The photo-detectors will be operated in a harsh radiation environment given by a total neutron fluence of \Neq{6} and 50\,Gy of ionising radiation.
Due to radiation, the \fdcr increases by several orders of magnitude over the lifetime of the detector.
In such an environment single photon detection and high noise rejection can, under certain conditions, still be achieved.

In our case, to cope with the radiation effect and the associated high \fdcr, four design parameters are crucial to allow for single photon detection: small active area 0.4\,\mm2, fast shaping and integration (25\,ns) adapted to the LHC bunch crossing rate, low correlated noise in particular low direct cross talk (see section~\ref{sec:time_const}) achieved by optical trenches between pixels and finally cooling to $-40$\degC. In total a reduction of \fdcr by a factor of 100 can be achieved compared to room temperature operation.

%% file: experimental_methods.tex
\section{Experimental methods overview}\label{exp_methods}

For the detector characterisation, two complementary setups are used.
The first one is used to collect the data for the pulse-shape analysis and PDE measurement. The data recorded are time dependent voltage pulses for the pulse-shape analysis and pulse frequencies for the PDE measurement obtained at different bias voltage settings. The second setup allows to record the IV characteristics used to measure \fdcr, \vbd, and the
quench resistor (\RQ).

\subsection{Pulse-shape time dependent measurement}

The pulse-shape measurements are performed inside an electromagnetic interference shielded enclosure to ensure low electronic noise. The detector is read out with a high bandwidth voltage preamplifier (2.5\,GHz, 20 or 40\,dBV\,\footnote{~FEMTO, HSA-X-2-20, HSA-X-2-40}) and a 1\,GHz oscilloscope (10\,GS/s)\,\footnote{~LeCroy, WAVERUNNER 104MXI}. The detector is mounted on a PCB avoiding wire pins or cables to minimise serial inductance and the associated ringing. The signal of the tested SiPMs is composed of two components, one fast and one slow.
The fast component has edges of less than 1\,ns and imposes to the mounting, cabling and read-out system to cope with high speed signals.
The bias voltage filtering is provided by a serial 1\,k$\rm{\Omega}$ resistor and two parallel ceramic capacitors (100nF and $1\,\rm{\upmu}$F) mounted close to the detector.
The box containing the device is equipped with a cooling system which allows to reach $-60$\degC reducing the \fdcr by typically more than a factor of 100 compared to ambient temperature operation.
The bias voltage source meter\,\footnote{~Keithley 2400} and the oscilloscope are controlled and read using Python routines.
The bias voltage current and a large number of waveforms of dark pulses are recorded at different bias voltages \vbias.
These are subsequently analysed using a program in C++ and ROOT~\cite{root_cern}.

\subsection{PDE measurement setup}

\begin{figure}[htbp]
\centering
\includegraphics[width=0.55\textwidth]{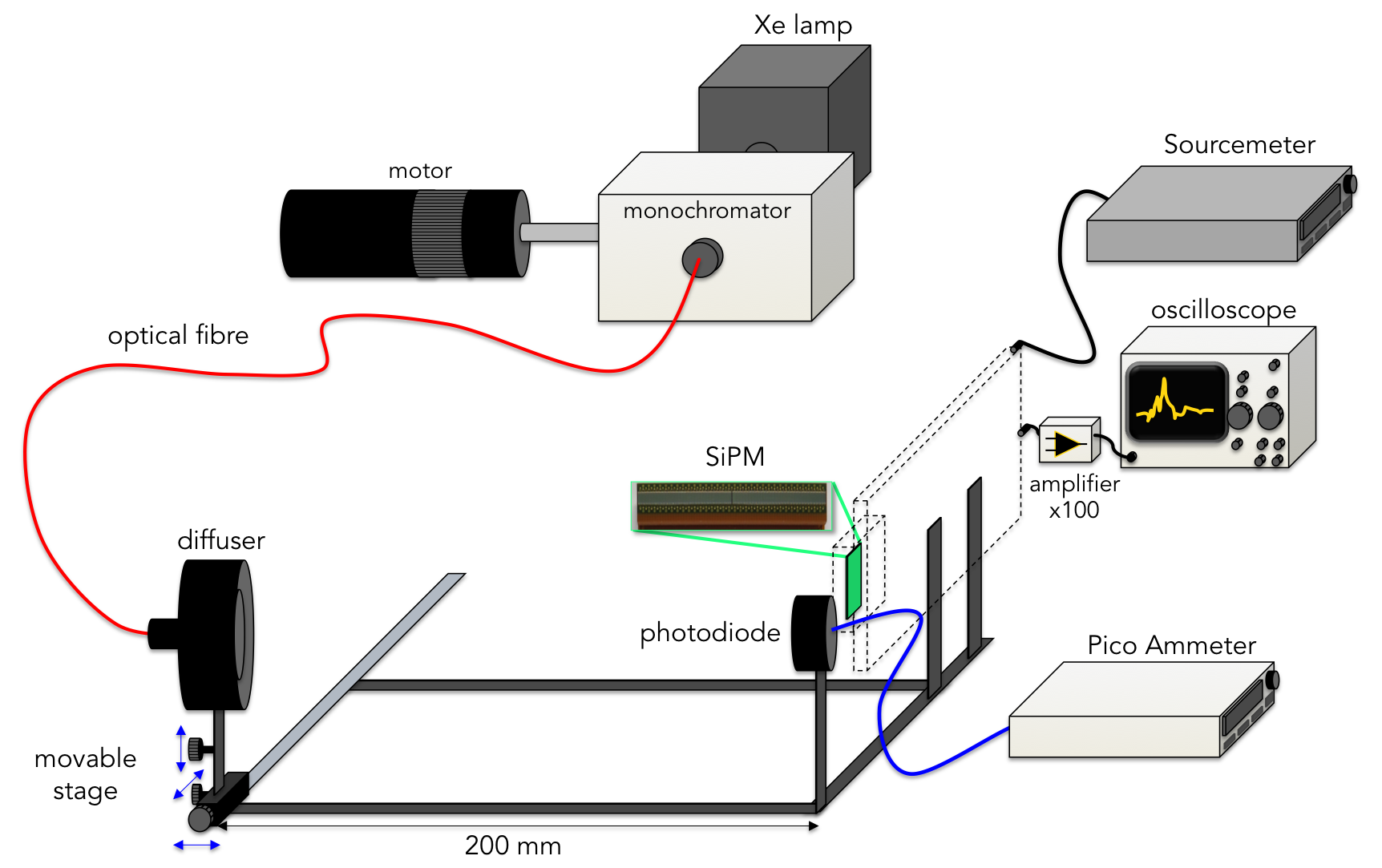}
\caption{Sketch of the PDE measurement setup.}
\label{fig:PDESetup}
\end{figure}

The setup used to measure the PDE as a function of wavelength and \dV, is shown in Fig.~\ref{fig:PDESetup}.
The light source is composed of a Xe lamp and a monochromator\,\footnote{~H10-UV Yvon Jobin} controlled by a step motor. The monochromator allows to select a narrow wavelength region $\pm$1\,nm of the Xe lamp spectrum (from 200\,nm to 750\,nm). The monochromator light output is coupled with a collimator\,\footnote{~Thorlabs fixed focus collimator F280SMA-A} into an optical fibre (550\,$\rm{\upmu}$m diameter).
The light beam is made homogeneous using a light diffuser and the separation of light source (diffuser) and detection plane (SiPM or photo-diode).
The calibration of the luminous power surface density \Plum is made by a calibrated photo-diode\,\footnote{~Newport 818-UV.}.

The photo-diode is sensitive in the range of 160 to 900\,nm and requires a relatively high light intensity compared to the SiPM.
Its current is read by a picoamper-meter\,\footnote{~Keithley 6485}.
The adjustment of the light intensity is made in such a way that the photo-diode receives sufficient optical power for the calibration while the SiPM not too much to avoid saturation.
The large sensitive area of the photo-diode (10.3\,mm diameter) is necessary to reach 1$\%$ precision for a minimal optical power of 100\,pW.
The relationship between  \Plum and the rate of incident photons, R, is: $\rm{P_{lum}= (R/A) \cdot hc/\lambda}$, where A is the sensitive area and $\rm{\lambda}$ the wavelength.
The luminous power density can also be calculated from the measured photocurrent using the relation $\rm{I= S \cdot A \cdot P_{lum}}$, where ${\rm{S}}$ is the radiant sensitivity, given by the manufacturer calibration curve.
Instead of the radiant sensitivity, the quantum efficiency, ${\rm QE= S \cdot hc/(\lambda} \cdot e)$, can be used. In our setup with a distance of 200\,mm, the surface of the photo-diode $\rm A=83.3$\,\mm2, the resulting \Plum is in the range of 1.25 to 5\,pW/mm$^2$ depending on the wavelength.
The maximum variation in the uniformity of the light intensity at the SiPM was found to 1\% over the active surface.
The light beam can be displaced on a \textit{xy}-positioning stage to find the peak luminosity.

\subsection{IV measurement setup}
The measurement of the IV characteristics is used to compute  \vbd and \fdcr as a function of \dV and \RQ.
The IV characteristics measurement does not require any high bandwidth data acquisition system and can therefore be  implemented with standard test equipment.
We use a source meter Keithley 2612B together with a multiplexer system\,\footnote{~Keithley 3706A-NFP} to automate the measurements for multichannel devices.
Be aware that the Keithley series 2600 is not a good choice for SiPM pulse-shape measurements but provides fast IV scan capability. This series introduces significant noise spikes from its own power supply into the test equipment which can be of the order of the 1 photon signal. To increase the precision of the \vbd measurement, a homogeneous light source can be applied without affecting the absolute value of the result. Examples of the recorded curves are given in Fig.~\ref{fig:RqfitandRq},\ref{fig:IVandDeriv}.

%% file: pulse_shape_analysis.tex
\section{Pulse shape analysis and correlated noise measurement}\label{noise_analysis}

The pulse-shape analysis allows to measure correlated noise probabilities as direct cross-talk, delayed cross talk and after-pulse as a function of \dV. It also gives a  statistical measurement for the most important time constants as long pulse component decay time, recovery time, delayed cross-talk mean lifetime and after-pulse mean lifetime.
The principle is to acquire on an oscilloscope a large number of dark pulses which are subsequently analysed off-line.
The characteristics are obtained as a function of $\rm{\Delta V=V_{bias}-V_{BD}}$.

One of the key features to enable these measurements is to have a sufficient signal-to-noise ratio (S/N) as already discussed in section~\ref{exp_methods} where the electronics circuit and amplifier are described.
This method was developed for small size SiPMs with an area of $0.4$\,\mm2 and  $\rm{f_{DCR}}<50$\,kHz ($\rm{\tau_{DCR}}=1/f_{DCR}=20$\,$\upmu$s) at room temperature. In this case, the small \fdcr allows to acquire data in a sampling window of $\rm{t_{sample}}=200$\,ns.
The probability of a second random dark pulse falling in the measurement period is $\rm{R_T = 200\,ns / 20}$\,$\upmu$s = 1\%.
To extend this method to larger devices with higher \fdcr we propose to perform the tests at low temperature. Changing the temperature naturally changes \fdcr but also, in a moderate way, AP. For certain devices, differences on the PDE were observed~\cite{Calouzol_FBK_cryo}. We have successfully performed the analysis with 1\,\mm2, 9\,\mm2 and 36\,\mm2 devices that have \fdcr as large as 1\,MHz at $\rm{T}=25$\degC which were cooled down to $-40$\degC. We suggest to operate the detector with a ratio $\rm{R_T}<2$\% to avoid significant contributions from random dark pulses in the sampling window. For the waveform analysis, we express all pulse amplitudes in the unit of ``photoelectron'' (PE), corresponding to a single pixel avalanche.

\subsection{Breakdown voltage measurement}\label{sec:vbd_pulse_shape}
In the following an overview of different methods to measure \vbd is given. We have observed significant differences (order of 400\,mV) between different methods. The differences observed can be explained by the method in case of the IV based measurement presented in section~\ref{sec:IV-VBD} or the data acquisition bandwidth for the amplitude based method presented here below. For the following discussion we propose to use the gain (charge) based \vbd value as a reference which fulfils the relation $\rm{G}\propto\Delta\rm{V}$.

\begin{figure}[htbp]
\centering
\includegraphics[width=0.55\textwidth]{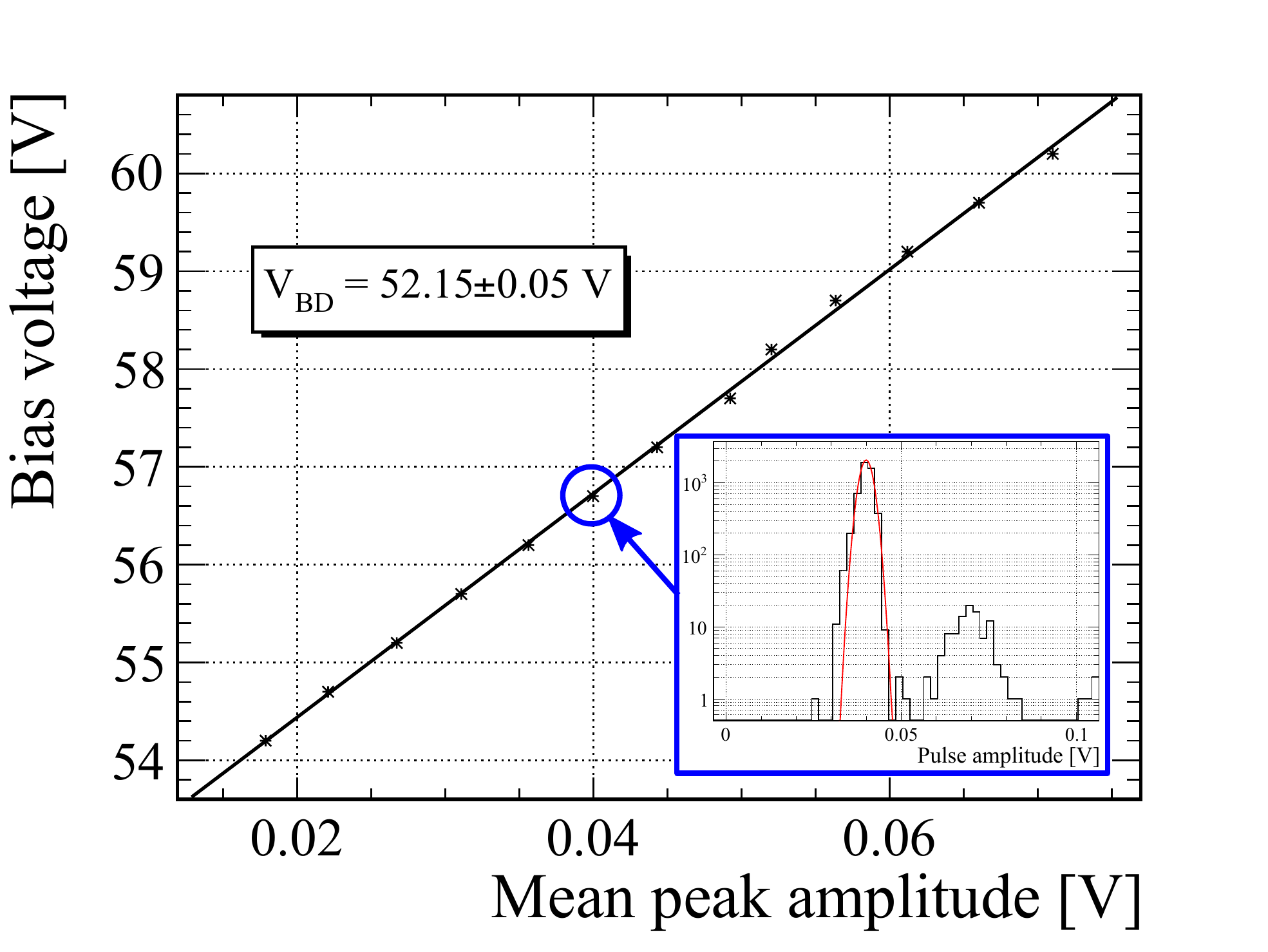}
\caption{Illustration of \vbd determination procedure. The 1\,PE amplitude $\rm A_{1\,PE}$ is measured with a fit of the primary pulse amplitude distribution with a Gaussian function (insert).}
\label{Vbd_calc_procedure}
\end{figure}

\paragraph{Pulse amplitude method}
The procedure to calculate \vbd in this way is illustrated in Fig.~\ref{Vbd_calc_procedure}.
For each bias voltage point, the 1\,PE pulse amplitude $\rm A_{1\,PE}$ is extracted from a Gaussian fit of the amplitude distribution of primary pulses.
The obtained values are fitted with a linear function and, given that G $\propto$ \dV, \vbd is calculated as the intersection of the fit with zero amplitude.
Since the duration of the measurement is sufficiently short to avoid any temperature changes, the \vbd obtained in this way can be used to calculate the \dV without further temperature compensation.
We call \vbdAmp the breakdown voltage obtained with this method, to differentiate with the \vbdI parameter obtained from the IV scan method in section~\ref{sec:iv}.
The uncertainty on \vbdAmp is typically between 10 and 50\,mV with a reproducibility better than 5\,mV.

\begin{figure}[htbp]
\centering
\includegraphics[width=0.55\textwidth]{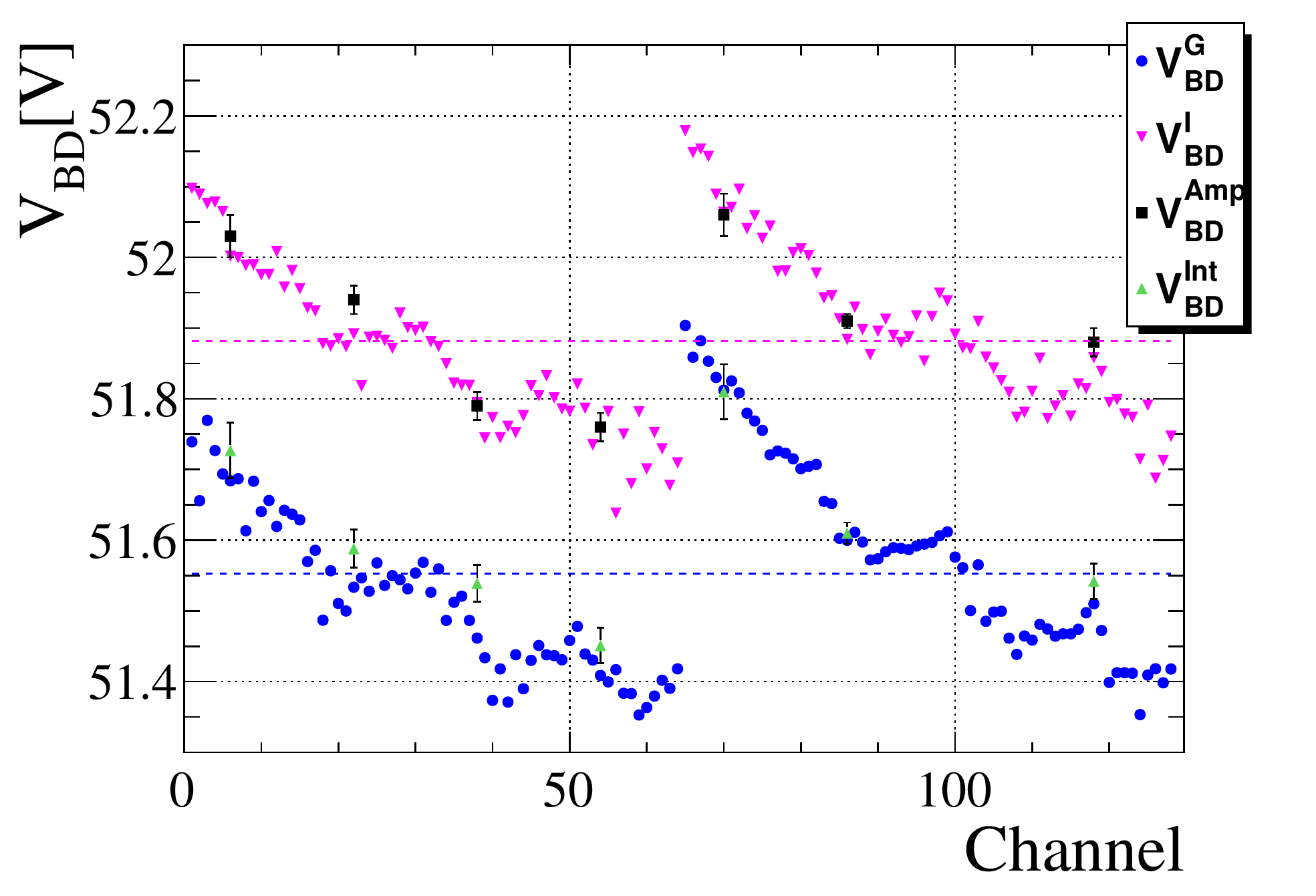}
\caption{Comparison of the results for \vbd obtained by four different methods for a 128-channel array device. \vbdG and \vbdInt obtained by the charge integration methods yield compatible results.}
\label{fig:VBD_IV}
\end{figure}

Comparing a large number of results, we observe a systematic difference of up to 400\,mV between \vbdAmp and \vbd measurements based on charge integration.
The linearity of $\rm A_{1\,PE}$ as a function of \vbias is maintained up to high \dV.
However, in the region from 0 to 0.3\,V, a deviation is observed which explains the offset in \vbd result.
The offset varies with the bandwidth of the read-out system.
Measurements taken with bandwidth limitation on the oscilloscope show that the \vbdAmp converges to the \vbd from charge integration as the fast pulse component is suppressed.
A comparison of the results for \vbd with different methods on the base of a 128-channel array is shown in Fig.~\ref{fig:VBD_IV}.

\paragraph{Charge integration of pulse method}
The recorded waveforms are used for a numerical integration leading to a charge integration method. The linear fit of the charge measured at each bias voltage point and its extrapolation to zero charge, is providing consistent results with other charge methods as discussed below. In Fig.~\ref{fig:VBD_IV} the obtained values with this method are called \vbdInt.

\paragraph{Charge integrator and amplifier ASIC}
The breakdown voltage can also be extracted from a low light amplitude spectrum using multichannel readout electronics. We denote the value found with this method as \vbdG. For the LHCb SciFi project, a data acquisition for 1024 channels based on VATA64~\cite{VATA64} ASICs with a linear response has been employed. The results obtained for \vbdG in this way are consistent within the expected systematic and statistical uncertainties with the values provided by the SiPM manufactures. The values of all 128 channels for one array are shown in Fig.~\ref{fig:VBD_IV}

\subsection{Correlated noise classification}
\begin{figure}[htbp]
\centering
\subfloat[Clean pulses \label{events_clean}]{\includegraphics[width=0.5\textwidth]{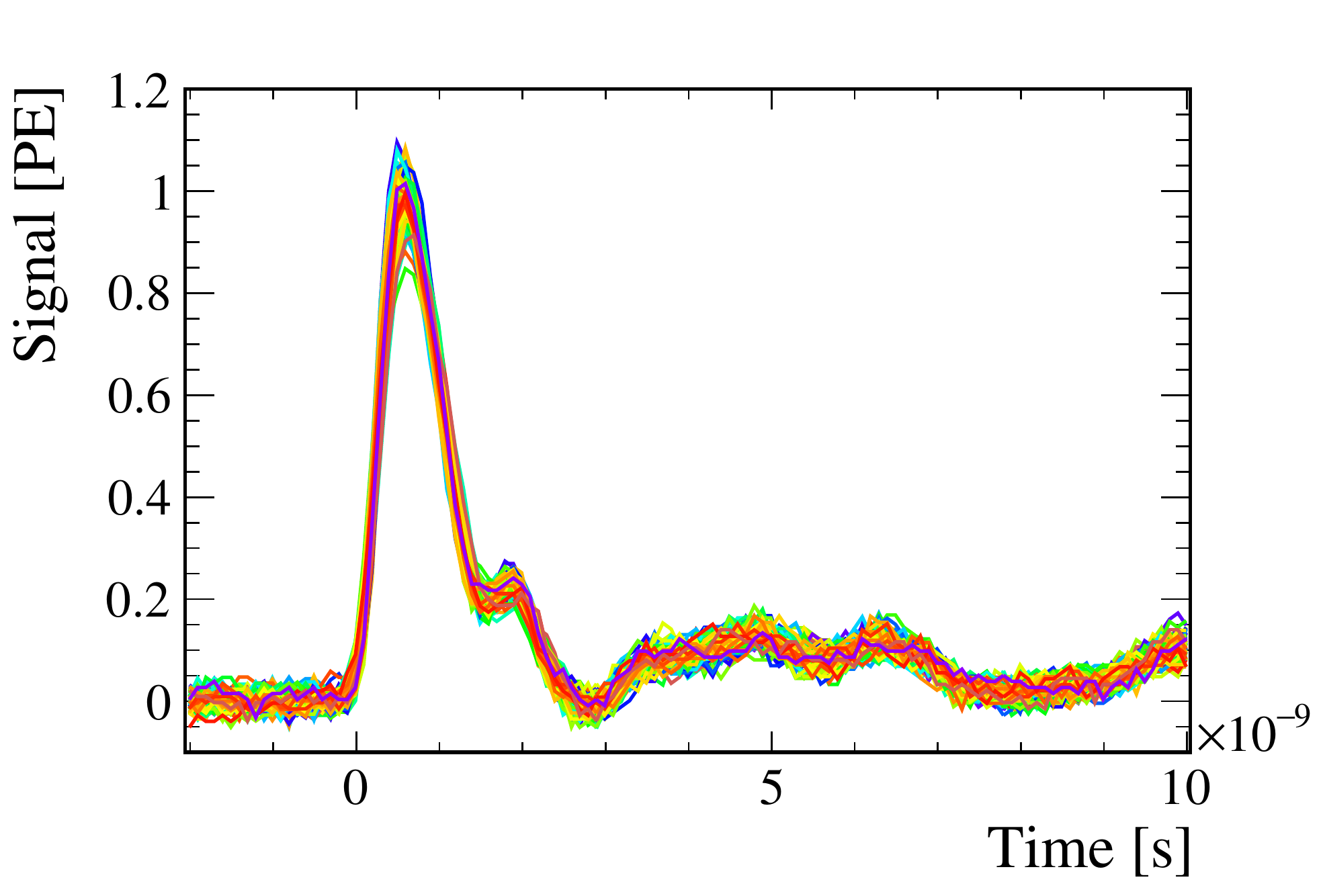}}
\subfloat[Pulses with DiXT \label{events_DiXT}]{\includegraphics[width=0.5\textwidth]{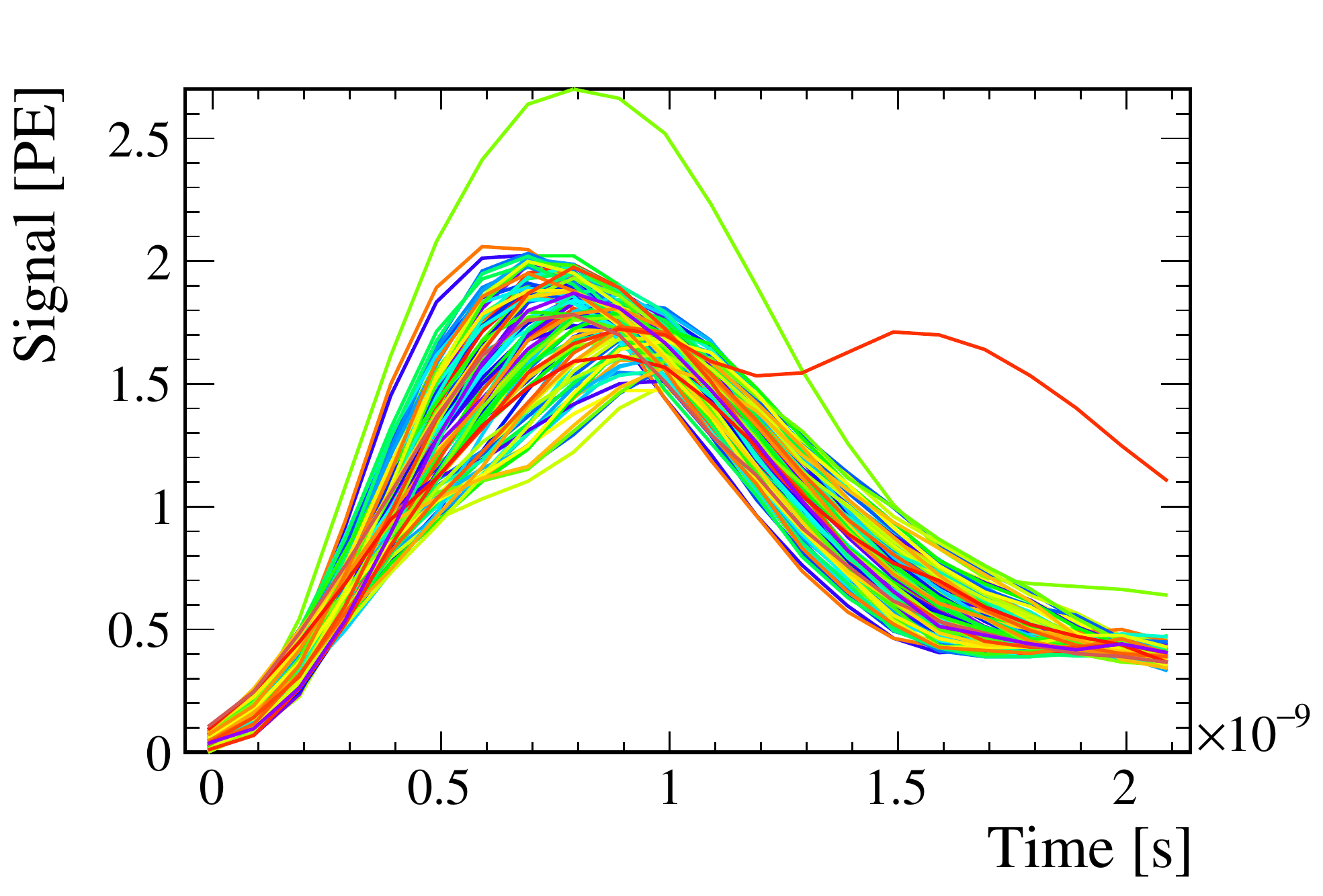}}\\
\subfloat[Pulses with delayed cross-talk \label{events_DeXT}]{\includegraphics[width=0.5\textwidth]{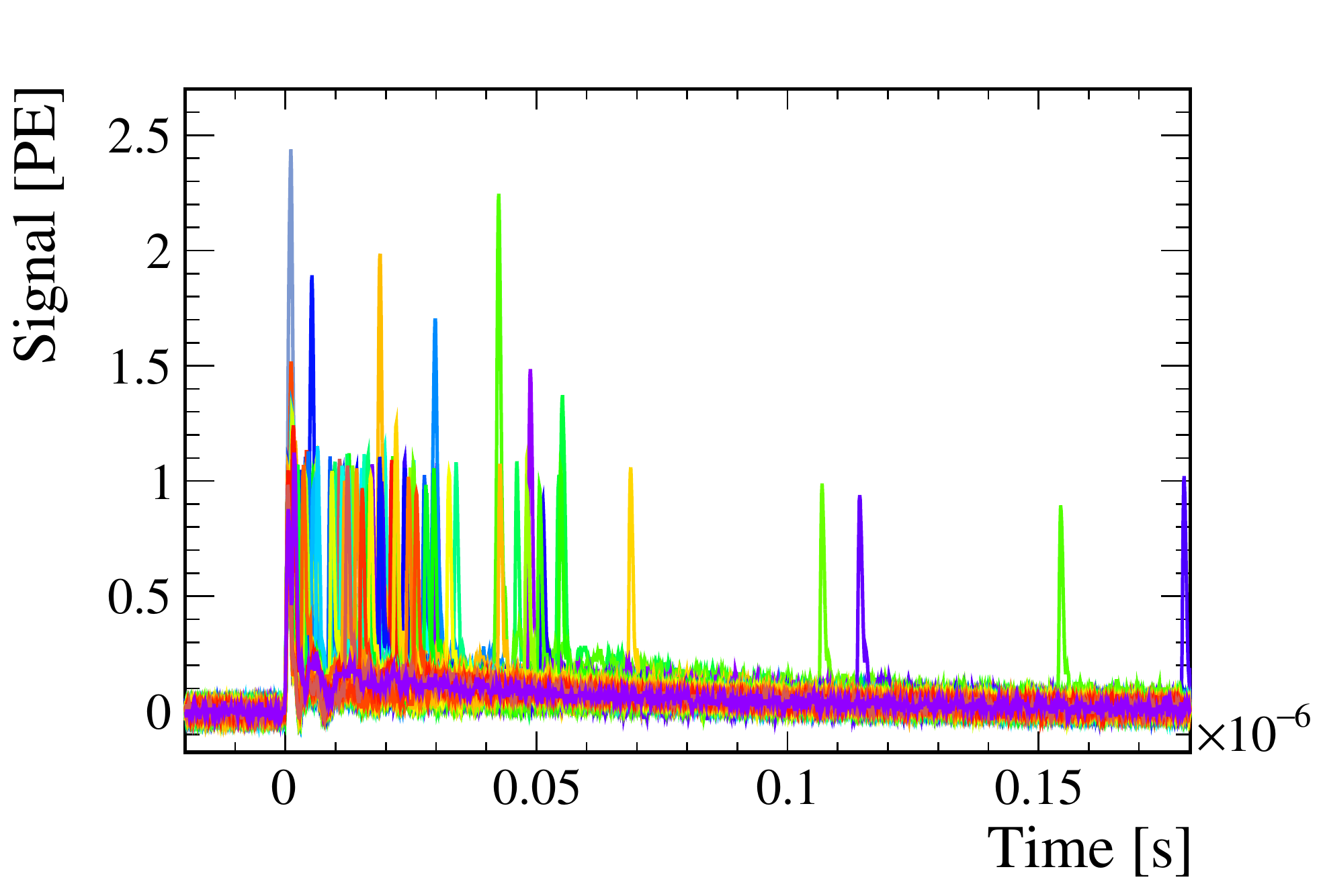}}
\subfloat[Pulses with after-pulses \label{events_AP}]{\includegraphics[width=0.5\textwidth]{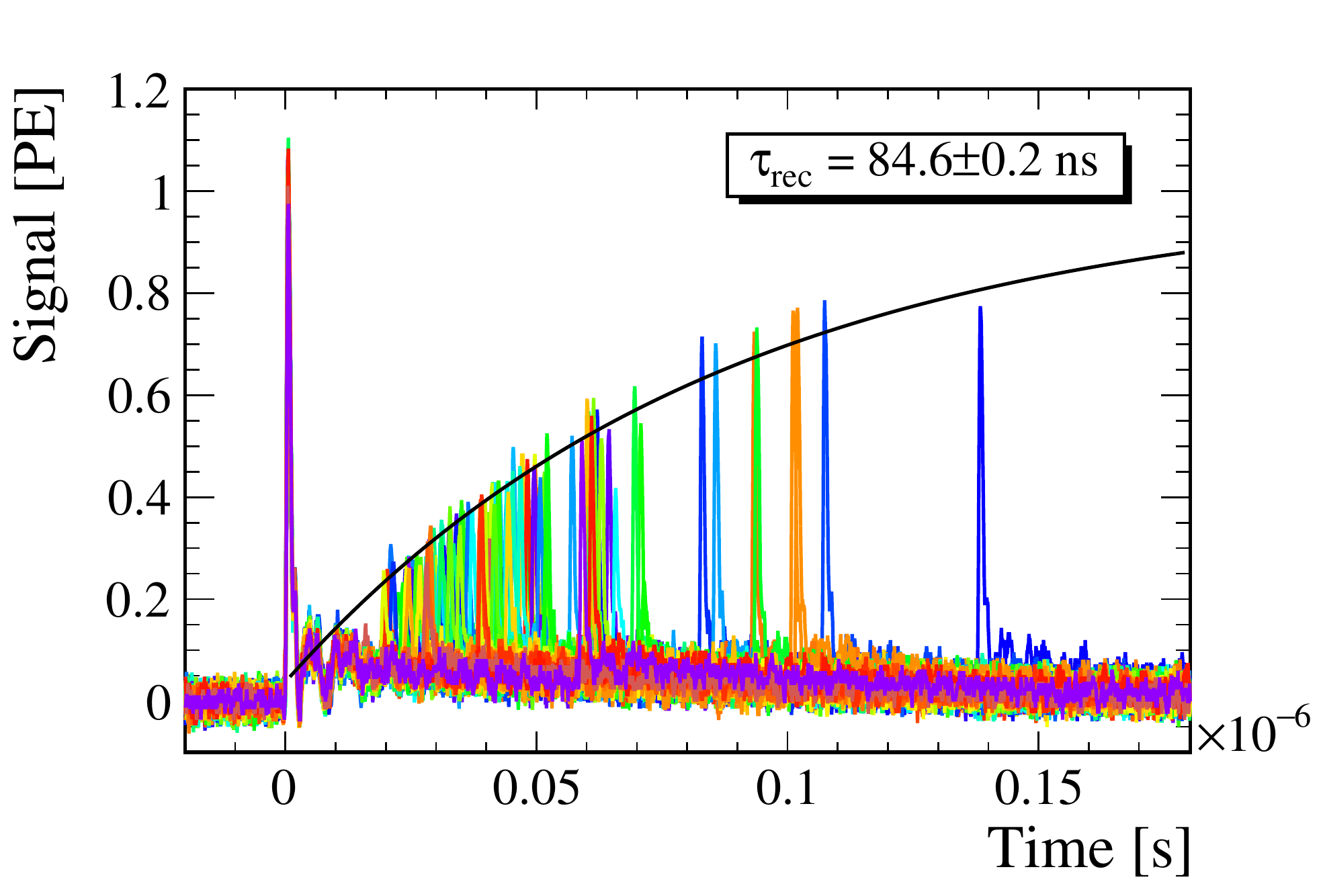}}
\caption{Examples of recorded waveforms. The figures show waveforms filtered out of 50k samples (maximum 100 waveforms printed out) in one graph after the waveform filtering algorithm. In (a) clean pulses with 1\,PE amplitude, (b) pulses with a DiXT and therefore $>1.17$\,PE amplitude, (c) delayed cross-talk, and (d) after-pulses and the exponential fit. In (c) one can observe delayed cross-talk which produce direct cross-talk (double height peaks). \label{events_overlay_zoom}}
\end{figure}

In a second step of the analysis, we identify noise pulses correlated with the primary dark pulse. Primary dark pulses without any related correlated noise are called clean events. The pulse shape of 100 clean waveforms superimposed are shown in Fig.~\ref{events_clean}. The filter algorithm can identify the following types of noise related to the primary dark pulses:
\begin{description}
\item[\textit{Direct optical pixel-to-pixel cross-talk (DiXT):}] Infrared photons are generated in a pixel when an avalanche is produced. They can hit a neighbouring pixel and create a secondary avalanche. This phenomenon is called optical pixel-to-pixel cross-talk and results in pulses produced simultaneously (or with small time delay of the order of a few 100\,ps) with respect to a primary pulse. The peak amplitude is 2\,PE for instantaneous cross-talk whereas it is between 1 and 2\,PE when the cross-talk is slightly delayed as it can be seen in Fig.~\ref{events_DiXT}. Higher amplitudes are possible due to multiple cross-talk.
\item[\textit{Delayed pixel-to-pixel cross-talk (DeXT):}] Pixel-to-pixel cross-talk is also observed with significantly larger delays compared to DiXT. This phenomenon can be explained by infrared photons absorbed deep in the bulk of the neighbouring pixels releasing electrons. These electrons are significantly delayed by the drift towards the avalanche region. In this case, secondary pulses have full amplitude (1\,PE) and their time of arrival is distributed in time as shown in Fig.~\ref{events_DeXT}.
\item[\textit{After-pulse (AP):}] In the pixel where a primary avalanche occurs, charge carriers can be trapped and released with some delay. The pulse amplitude depends on the recovery state of the pixel and its value ranges from 0 and 1\,PE as shown in Fig.~\ref{events_AP}. The time of arrival of AP is distributed over a few 100\,ns. A much smaller contribution with longer time constant is also observed which is for the noise analysis in our case neglected. A fit of the AP amplitude as a function of arrival time is used to extract the exponential recovery time.
\end{description}

\paragraph{Amplitude thresholds and time windows}
To distinguish DiXT, DeXT and AP, the filter algorithm uses amplitude thresholds and time windows. The amplitude thresholds are set after the calibration in PE units, done for every \vbias value, as discussed in section~\ref{sec:vbd_pulse_shape}. The time window is set relative to the rising edge of the primary dark count pulse. Amplitude thresholds and time windows may be adjusted differently for different SiPM types to adapt to different pulse-shapes or electronic noise levels. Typical values are 1.17\,PE in a time window 0 to 1\,ns for DiXT, 0.85\,PE for times $>1$\,ns for DeXT and 0.25\,PE for times $>10$\,ns for AP\footnote{To compare the AP probability with the manufacturer specification, the threshold is set 0.5\,PE}. The trigger amplitude and voltage scale of the oscilloscope are adjusted for each \vbias to avoid saturation or triggering on electronic noise. The time window is $[-20,180]$\,ns around the rising edge of the dark pulse. Due to the serial inductance of the SiPM package some ringing is observed after the fast transient of the fast component of the signal. A 10\,ns region is discarded for the after-pulse detection to avoid misidentification.

\paragraph{Pulse filtering}\label{par:Pulse filtering}
The pulses are classified by the criteria described above. The data allows to measure the occurrence probabilities of the different correlated noise. Each waveform may contain only the primary noise pulse, or a single or several correlated noise pulses.
Every waveform containing correlated noise peaks is classified as DiXT, DeXT or AP, according to the nature of the first correlated noise peak. If for a single recorded waveform more than one correlated noise peak is detected, the additional peaks are counted as \textit{higher order} correlated noise. For high correlated noise probabilities, the higher order noise is significant. The probability of correlated noise of the kind $X$ (DiXT, DeXT or AP) is the ratio of the number of classified waveforms ($\rm{N_{X}}$) to the total number of waveforms ($\rm{N_{ev}}$): $\rm{p_{X}=N_{X}/N_{ev}}$.

\subsection{SiPM electrical model}\label{sec:model}
Using the SiPM equivalent circuit model introduced by~\cite{Corsi_Asic_development_for_SiPM_readout}, the time constants and the model parameters can be calculated and verified. The two fast time constants, the pulse rise time and the fast decay time constant are out of reach with the present data acquisition system. The observed rise and fall times are dominated by the bandwidth limitation of the oscilloscope. On the other hand, as will be shown in section~\ref{sec:time_const} excellent precision has been achieved on the measurement of the gain (G), the recovery time constant (\trec) and the long component time constant (\tlong).
From measurements, we extract the values for the pixel diode capacitance $\rm{C_d}$ and the capacitance parallel to the quench resistor $\rm{C_Q}$.
The sum $\rm{C_d +C_Q}$ can be calculated from the gain given the relation $\rm{G}\cdot \ensuremath{e} = \dV \cdot (C_d +C_Q)$, and from the recovery time given $\rm{\tau_{rec}} = \rm{R_Q}\cdot (\rm{C_d+C_Q})$.
The two obtained values are compatible.

In contrast to the perfect agreement with the model for \trec, the experimental data for the long time constant \tlong follows the unexpected relationship $\rm{\tau_{long} = R_Q\cdot C_d}$, where from the electrical model expectation, the relation for \tlong and \trec should be identical. This difference was observed on several types (Hamamatsu and KETEK) of detectors with large fast components as used for the LHCb SciFi project.

\subsection{Illustration of the method with H2017 detectors}\label{sec:time_const}
\begin{figure}[htbp]
\centering
\subfloat[Primary correlated noise composition \label{prim_corr_summary}]{\includegraphics[width=0.5\textwidth]{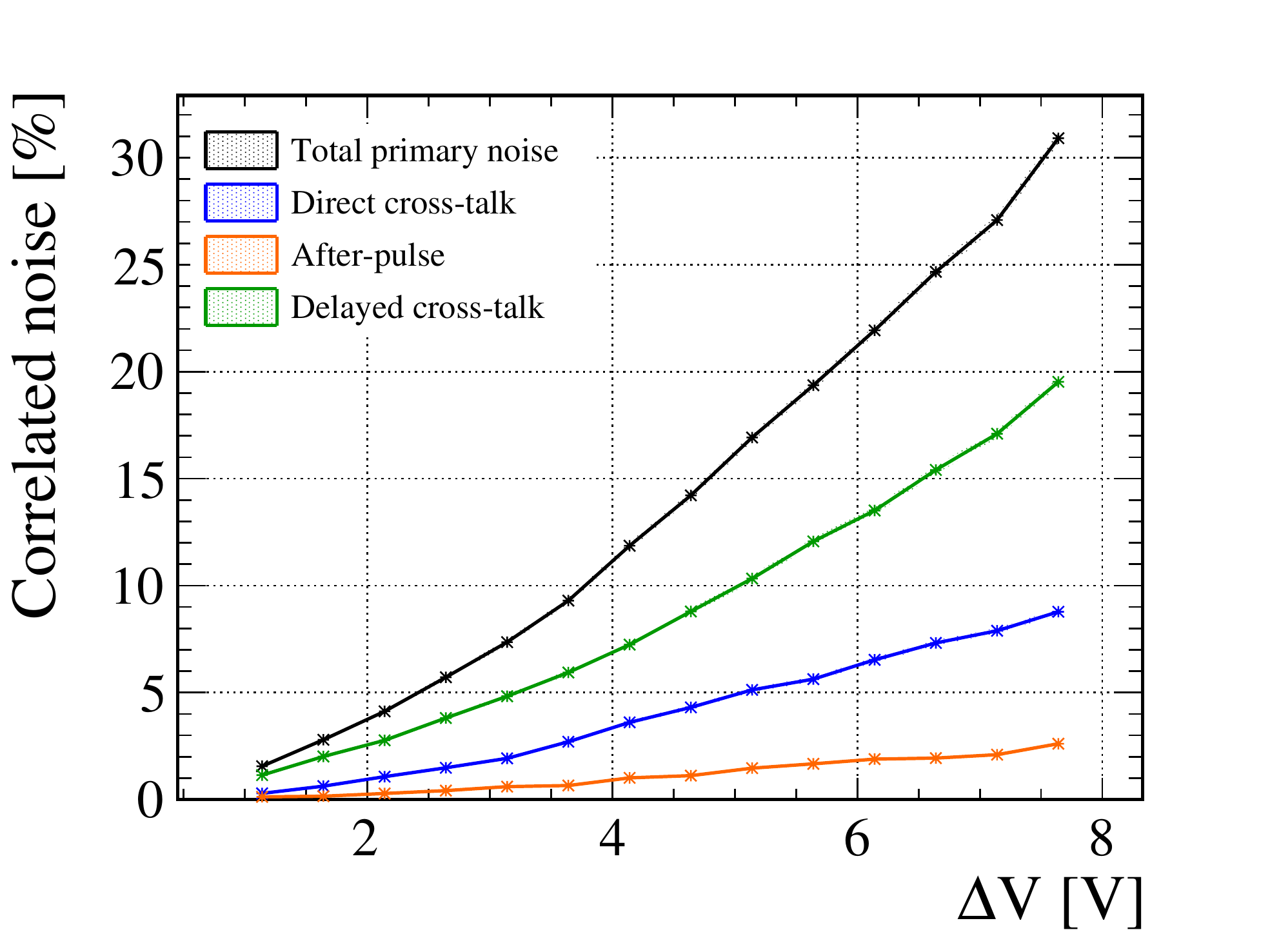}} 
\subfloat[Primary, higher order and DCR contributions \label{prim_vs_sec_noise}]{\includegraphics[width=0.5\textwidth]{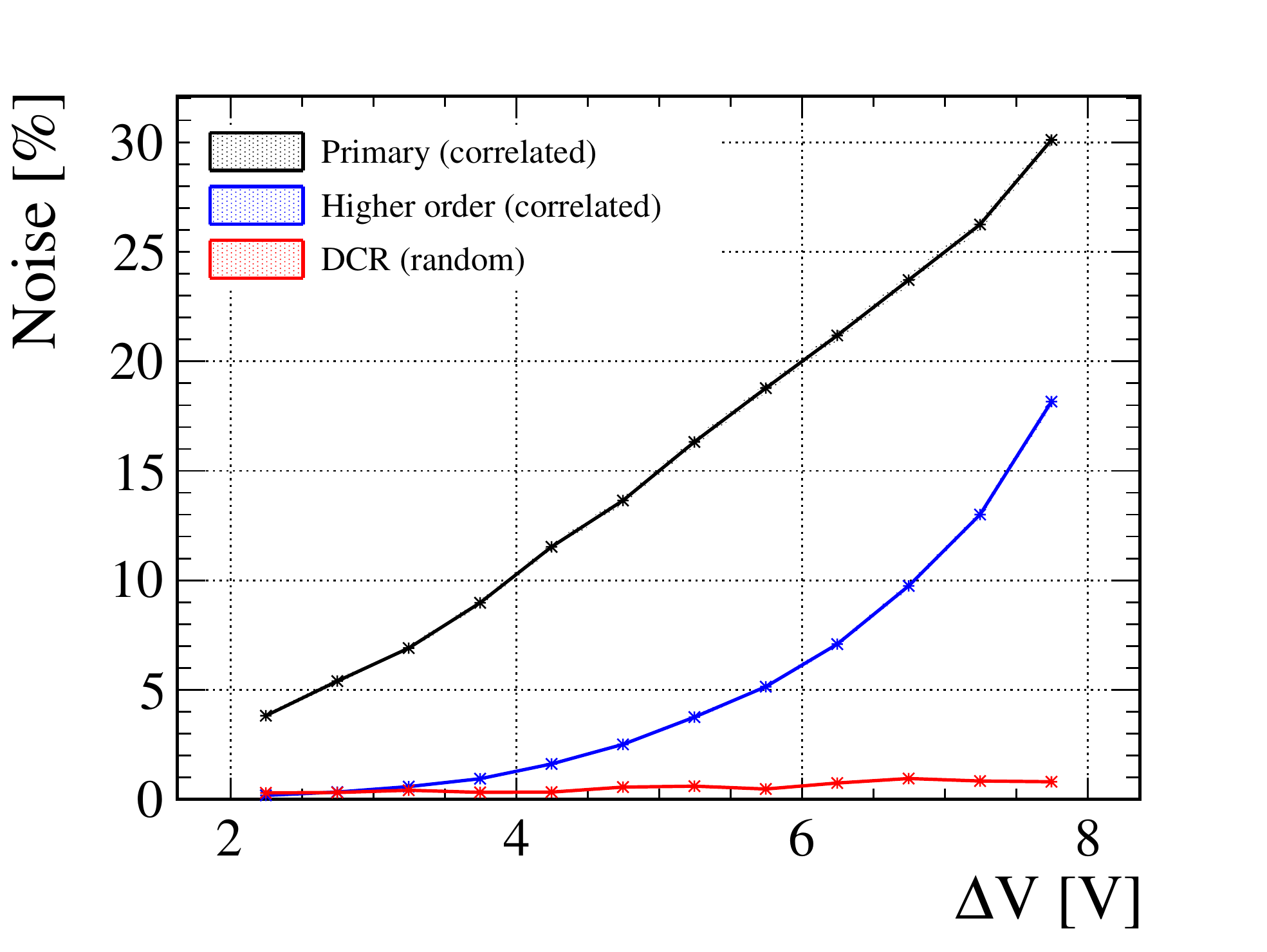}}
\caption{Correlated noise probabilities for an H2017 detector as a function of \dV (a) split into the different contribution types and (b) total primary and higher order correlated noise. \label{primary_secondary_corr_noise}}
\end{figure}

\begin{figure}[htbp]
\centering
\includegraphics[width=0.55\textwidth]{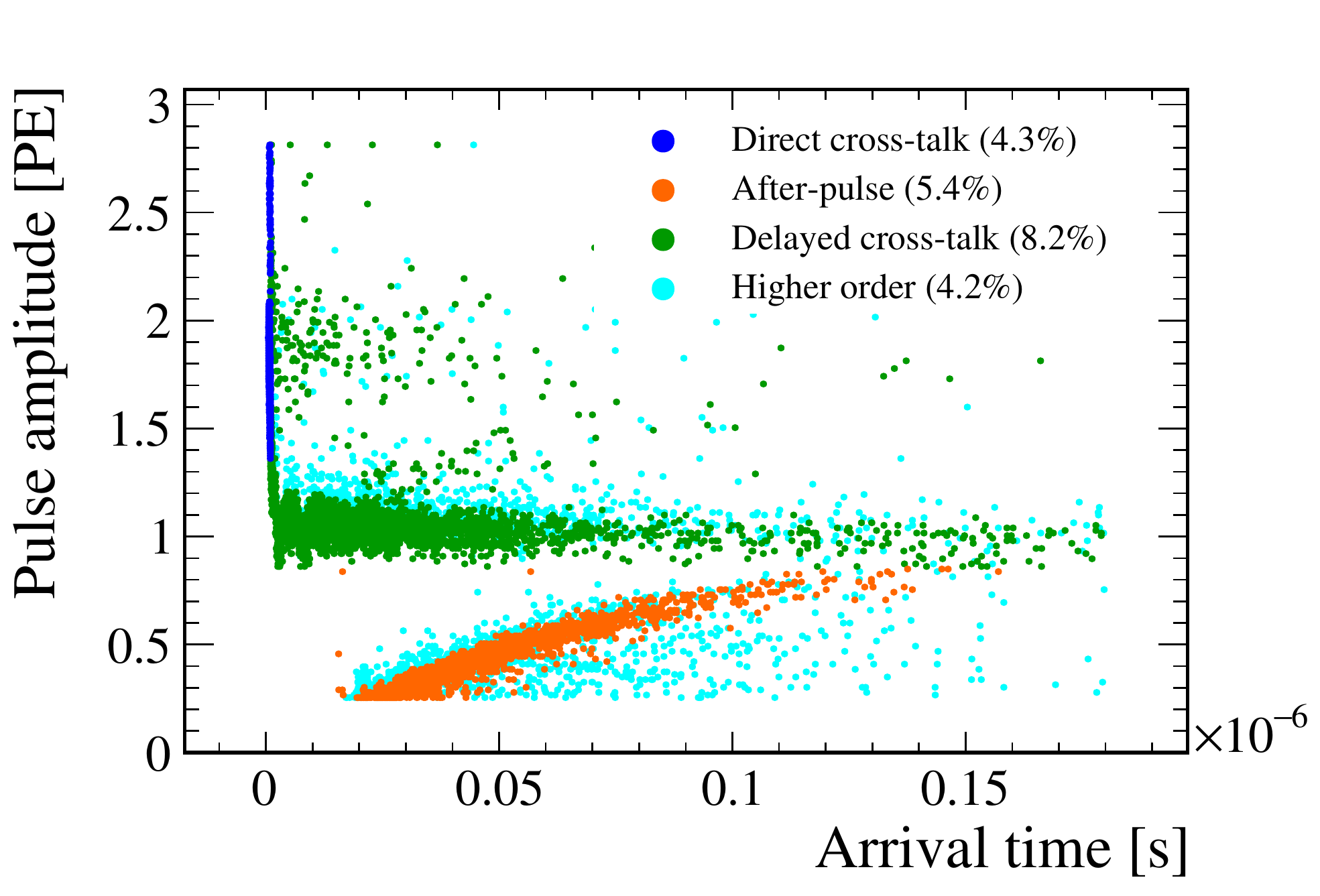}
\caption{Amplitude of classified correlated noise pulses as a function of arrival time.}
\label{amp_vs_time}
\end{figure}

\begin{figure}[htbp]
\centering
\includegraphics[width=0.55\textwidth]{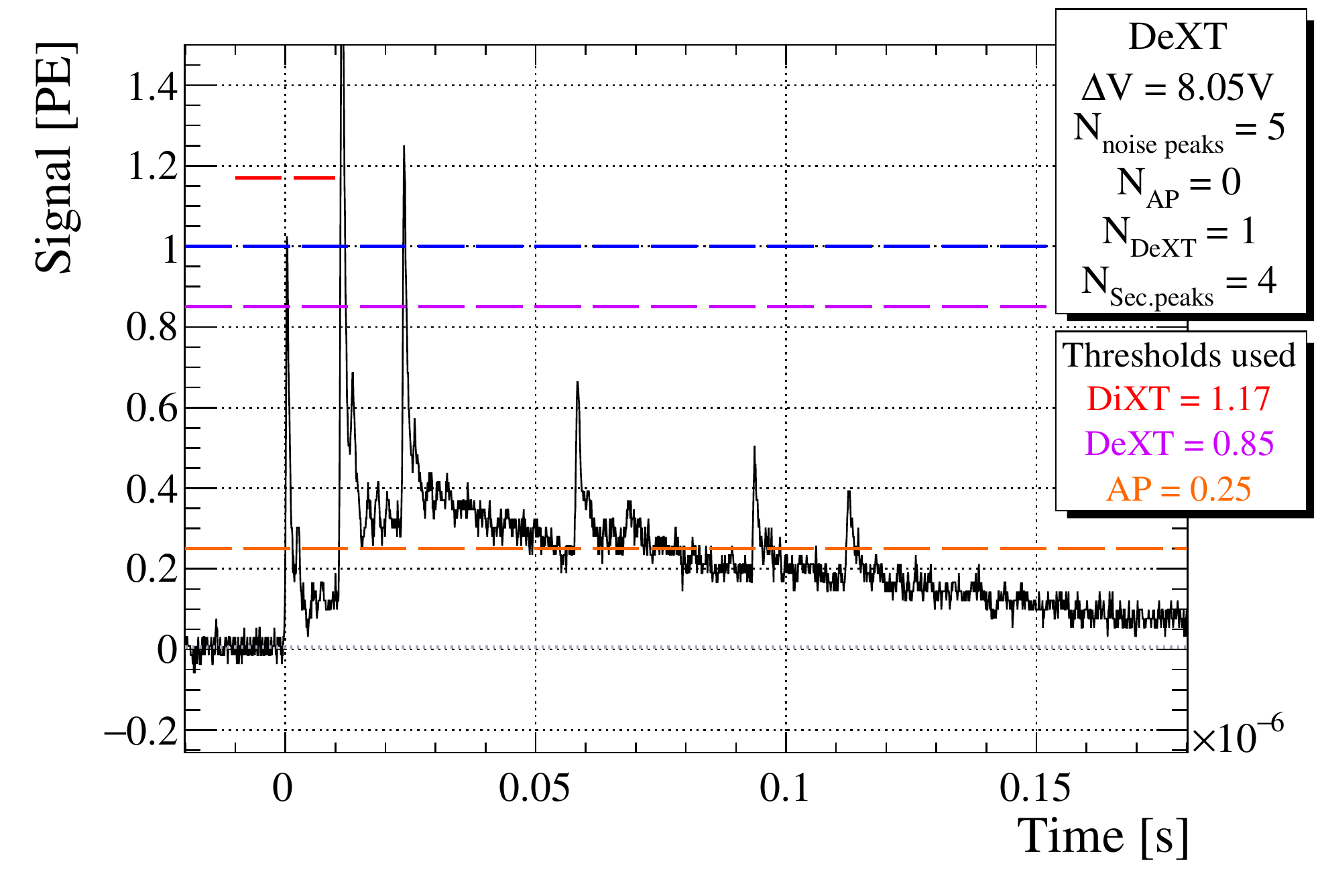}
\caption{Pulse waveform classified as delayed cross-talk with additional higher order correlated pulses recorded at high \dV (H2017 at \dVeq{8.0}).}
\label{nasty_event_display}
\end{figure}

The correlated noise probabilities for an H2017 detector is shown in Fig.~\ref{prim_corr_summary}. The detector shows significant contributions from all three correlated noise sources DiXT, DeXT and AP.
Higher order noise (noise of noise) represents an important contribution if the total primary correlated noise exceeds $20$\%, as shown in Fig.~\ref{prim_vs_sec_noise}.
For illustration, Fig.~\ref{amp_vs_time} shows the amplitude versus arrival time distribution of correlated pulses. In this view, one can observe the DiXT pulses composed of single and multiple simultaneous DiXT (up to 3\,PE is reached), DiXT from DeXTs and APs with amplitudes between $1-2$\,PE can be observed. As shown in Fig.~\ref{nasty_event_display}, the superposition of correlated pulses (higher order correlated pulses) can lead to large signal pulses.

\begin{figure}[htbp]
\centering
\subfloat[AP \label{fit_APtime}]{\includegraphics[width=0.5\textwidth]{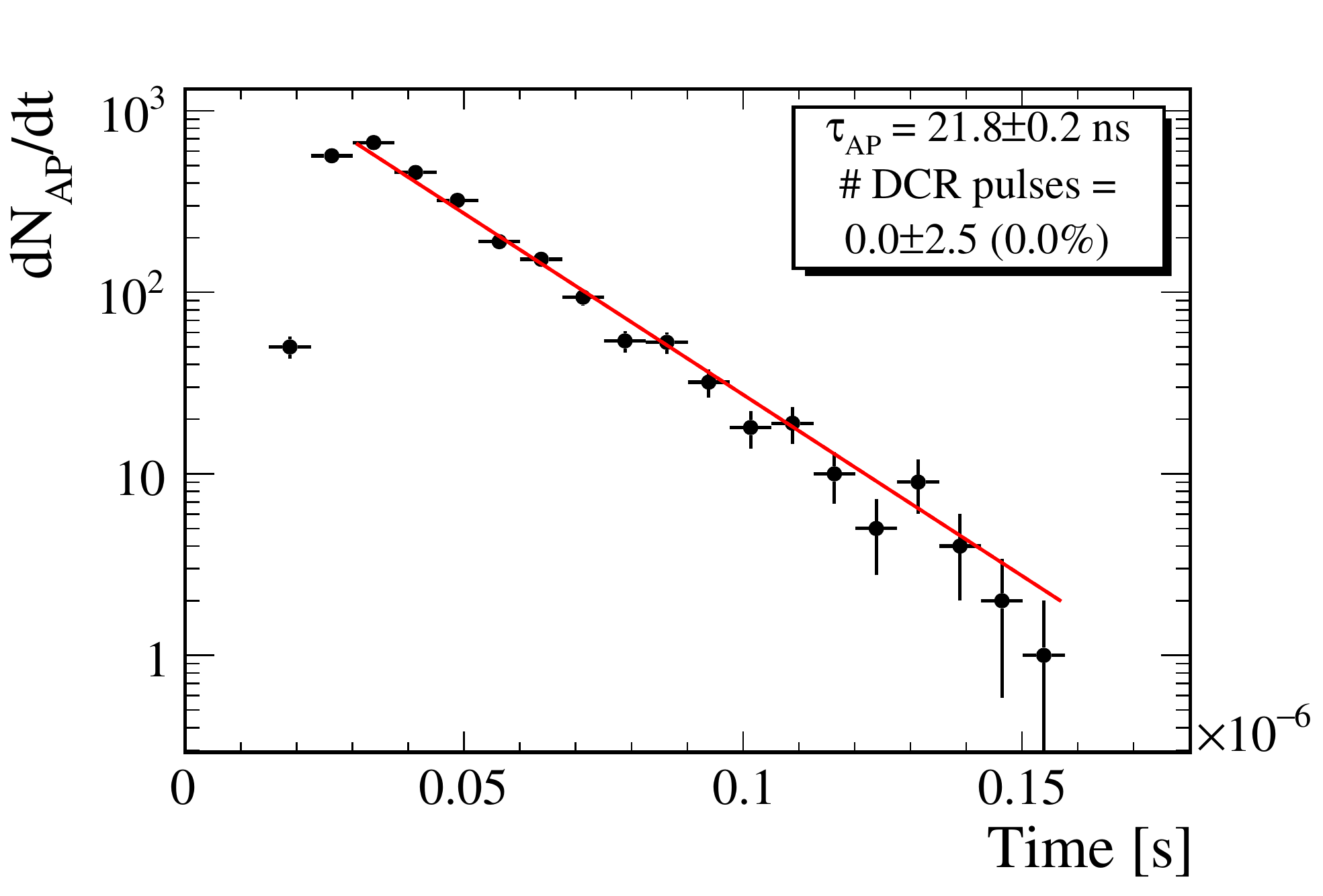}} 
\subfloat[DeXT \label{fit_DeXTtime}]{\includegraphics[width=0.5\textwidth]{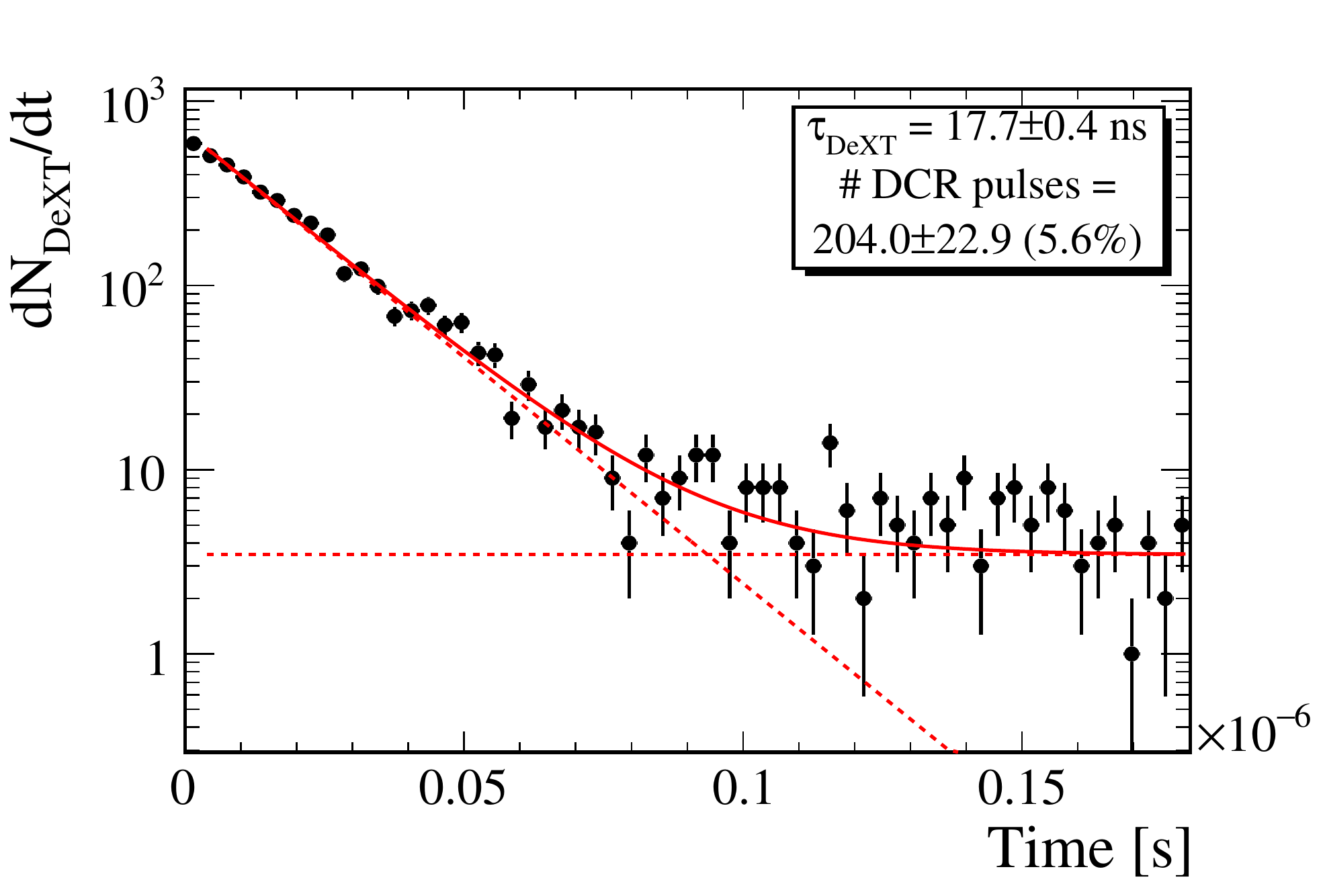}}
\caption{Mean lifetime determination and dark pulse contribution estimation from the arrival time distribution. \label{fit_mean_lifetime}}
\end{figure}

\paragraph{AP, DeXT and random dark pulses misidentification}
After a time of $\rm{2\cdot\tau_{rec}}$, the AP pulse amplitude reaches the DeXT amplitude threshold of 0.85\,PE. At this point the pulses are indistinguishable and classified as DeXT. This effect is however small since the number of APs decreases exponentially with time as shown in Figs.~\ref{amp_vs_time} and \ref{fit_APtime}.

For detectors with high \fdcr, random dark pulses are likely to occur in the sampling window of 180\,ns after the primary dark pulse.
The filter algorithm  accounts these pulses as DeXT.
The contribution of DCR to the DeXT can however be evaluated from the distribution of arrival time as shown in Fig.~\ref{fit_DeXTtime}.
In this way, the fraction of all detected delayed pulses attributed to DCR is monitored (see Fig.~\ref{prim_vs_sec_noise}).
In this particular example, it is in the order of 1\%, as expected from the calculation of $\rm{R_T}$ at the beginning of this section.

\begin{figure}[htbp]
\centering
\subfloat[Long decay time constant fit in log scale  \label{fit_longtau}]{\includegraphics[width=0.5\textwidth]{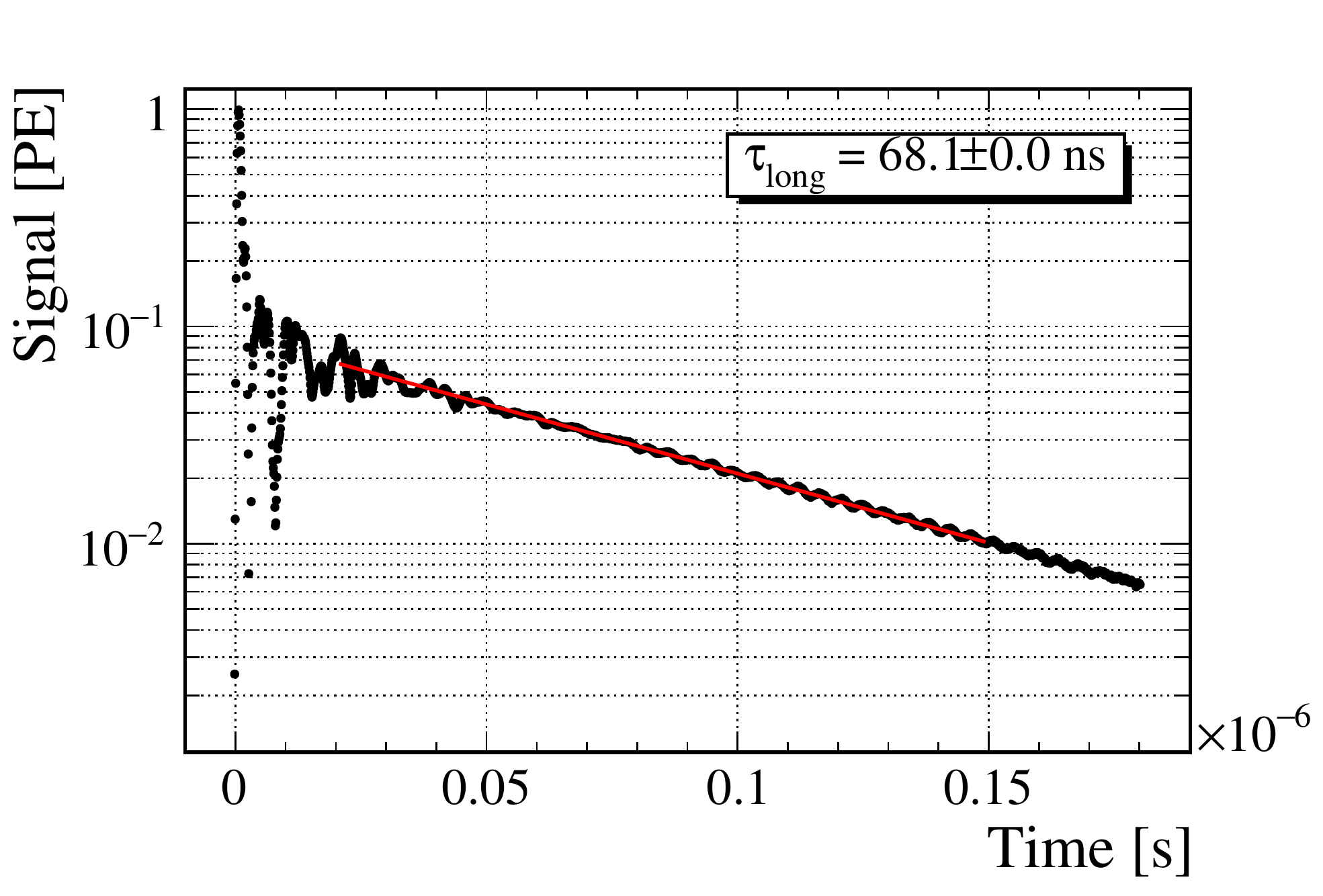}} 
\subfloat[Recovery time fit \label{fit_recovery}]{\includegraphics[width=0.5\textwidth]{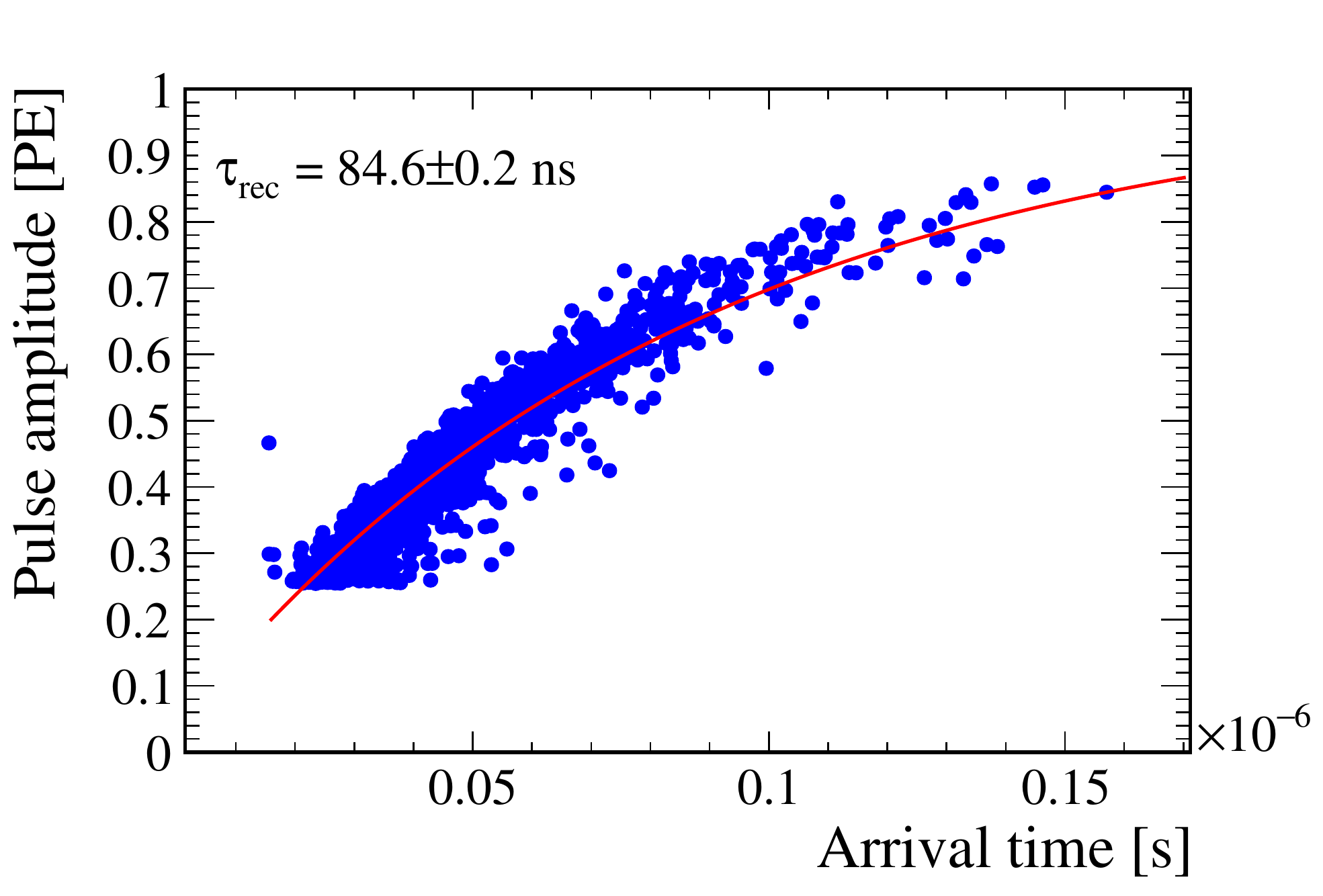}}
\caption{Determination of pulse-shape long decay time constant and recovery time constant for an H2017 detector. \label{time_constant_determination}}
\end{figure}

\paragraph{Long component time constant}
For clean waveforms, a double exponential function fits the fast and slow components of the falling edge of the pulse, with time constants \tshort and \tlong. For devices with small $\rm{C_Q}$ and therefore a small fast component, the slow and the fast components become indistinguishable. The fast component has a time constant <1\,ns and is dominated by the bandwidth of the read-out (in this case the oscilloscope). In Fig.~\ref{fit_longtau} the fit of \tlong is illustrated on the clean pulse waveforms.

\begin{figure}[htbp]
\centering
\includegraphics[width=0.55\textwidth]{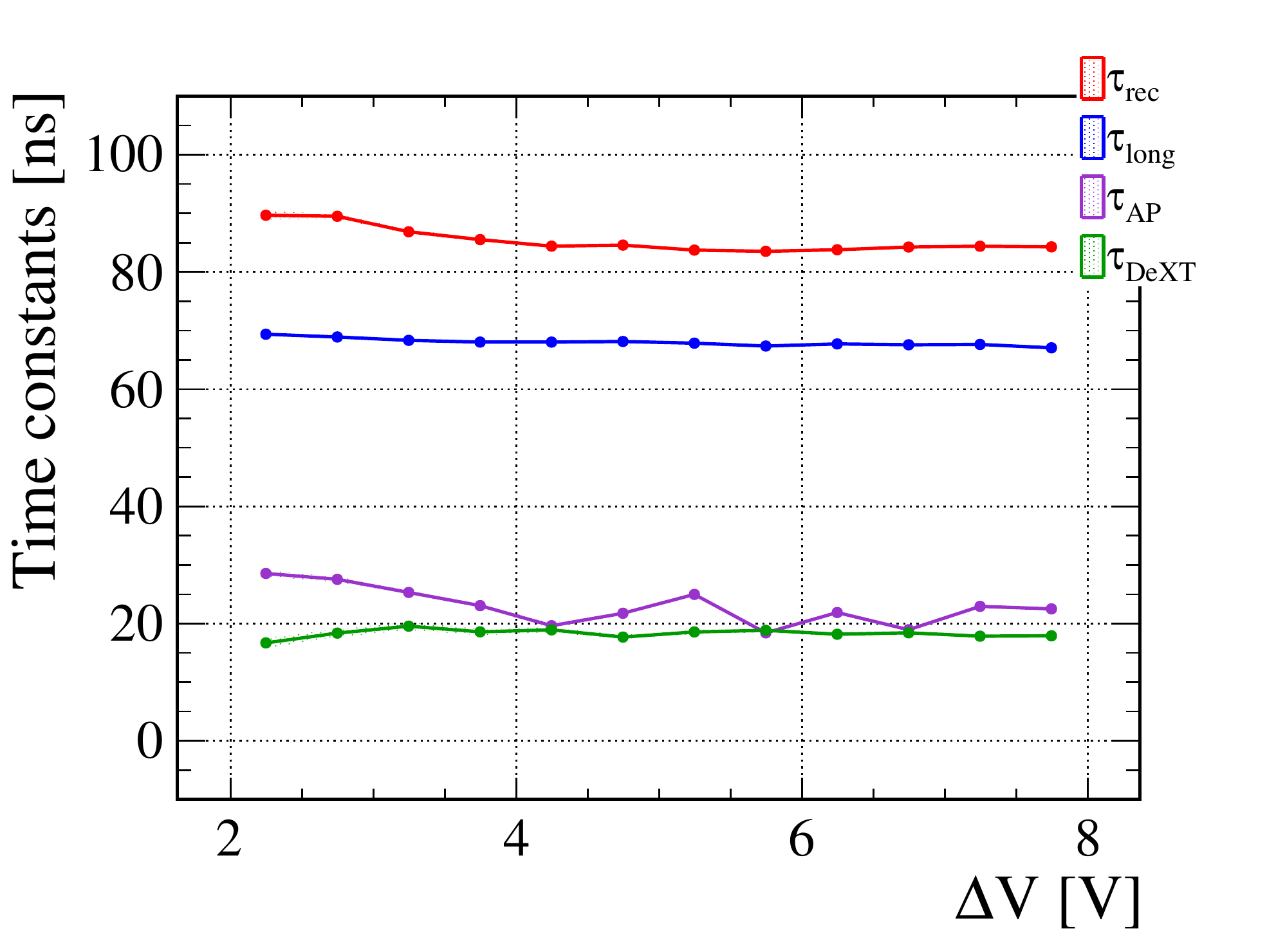}
\caption{Fit results of the time constants for different \dV.}
\label{fig:time_constant_plot}
\end{figure}

\paragraph{Recovery time constant}
The pixel recovery time constant \trec is measured from AP waveforms as shown in Fig.~\ref{events_AP} and \ref{fit_recovery}. The amplitude is assumed to recover with time as $1-e^{\rm{-\frac{t-t_0}{\tau_{rec}}}}$, where $\rm{t_0}$ is the  time needed for the recovery to start. A correction to the amplitude of the AP is introduced to compensate for the slow component of the primary pulse. The fit function is:
\begin{equation}
\rm{A(t)=A_{1\,PE}}\cdot\left(1-e^{-{\frac{\rm{t-t_0}}{\tau_{rec}}}}\right)+\rm{A_{slow}}\cdot e^{-{\frac{\rm{t}}{\tau_{long}}}},
\end{equation}
where $\rm A_{slow}$ is the amplitude of the slow component of the SiPM pulse.
The two parameters $\rm A_{slow}$ and \tlong are fixed by the pulse-shape fit results and $t_0$ is fixed to 4\,ns\,\footnote{~The value for $\rm{t_0}$ has been fixed to the average observed value for robustness, it can be left free for verification purpose.}. In Fig.~\ref{fit_recovery} the fit of the recovery time is illustrated. Consistent results within the expected uncertainties for \trec over a large operation range were obtained as shown in Fig.~\ref{fig:time_constant_plot}.

\paragraph{Time constants and the SiPM model for H2017}
The obtained results for the time parameters of the signal waveform allow to calculate the electrical model parameters.
With \tlongeq{67}, \treceq{84} and \RQeq{503} obtained from the IV measurement (section~\ref{sec:iv}), the values for the capacitors of the model can be calculated. With the relation for \tlong and \trec of section~\ref{sec:model}, we compute $\rm{C_d = 136}$\,fF and $\rm{C_Q = 34}$\,fF. With the two capacitor values the detector gain can be calculated as  $\rm{G/\Delta V =(C_d +C_Q )/e=}1.05\cdot10^6$\, $V^{-1}$ and compared with the gain obtained from pulse frequency and current measurement presented in section~\ref{sec:gain} where a value of $1.02\cdot10^6$\, $V^{-1}$ was found.

\paragraph{AP and DeXT mean lifetimes}\label{ap_lifetimes}
The AP filtering procedure allows to calculate the AP mean lifetime \tap with an exponential fit of the arrival time distribution as shown in Fig.~\ref{fit_APtime}. For the H2017 detector, the value of \tap is constant in the observed temperature range (-40\degC to +25\degC) and for different \dV, as seen in Fig.~\ref{fig:time_constant_plot}. Using \tap and \trec, one can calculate the fraction of missed APs due to the amplitude and time thresholds applied. For a threshold of 0.5\,PE, the fraction of missed APs is 93\%.

An exponential function is also used to describe the arrival time of DeXT pulses and allows to obtain $\rm{\tau_{DeXT}}$ as shown in Fig.~\ref{fit_DeXTtime}. The offset from the baseline is measured and attributed to random dark pulses present during the measurement. In Fig.~\ref{fit_DeXTtime}, 204 delayed pulses are attributed to dark pulses for a total of 50k waveforms. This corresponds to 5.6\% of the detected DeXT pulses.

\subsection{Pulse shape analysis for different devices}

\begin{figure}[htbp]
\centering
\includegraphics[width=0.55\textwidth]{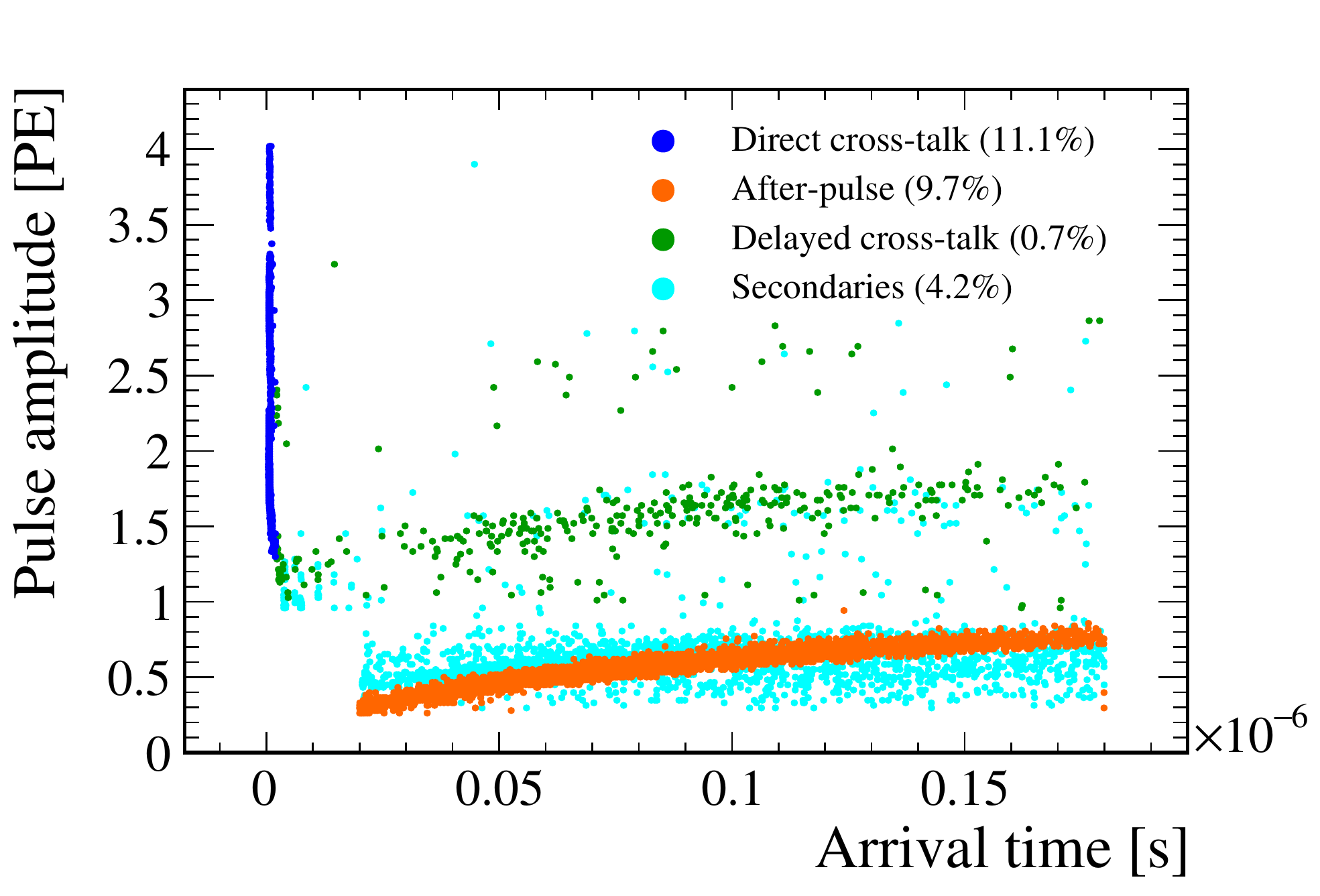}
\caption{Amplitude of classified correlated noise pulses as a function of arrival time for a KETEK prototype at \dVeq{5.2}.}
\label{ketek_amp_vs_time}
\end{figure}

The pulse-shape analysis was applied to SiPMs of several suppliers and technologies with different total area and pixel size. As an example, a KETEK prototype\,\footnote{~$62\times 62$\,$\upmu$m$^2$ pixels with thick EPI layer} tested from \dVeq{1\rm{~to~}5} showed almost zero DeXT and very small AP probability. Figure~\ref{ketek_amp_vs_time} shows the amplitude versus arrival time diagram of correlated noise at high \dV. As it can be seen, the pulses identified as DeXT are in fact APs with a DiXT and, consequently, the DeXT probability scales as $\rm{ p_{DeXT} \sim p_{DiXT}\cdot p_{AP} \simeq 1\%}$.

For single channel devices with smaller pixel size (and therefore smaller gain) from FBK, SensL and Hamamatsu\,\footnote{~FBK RGB-HD-SiPM $25 \times 25$\,$\upmu$m$^2$ pixels, SensL MicroES-SMA-30035-TSV-E32, Hamamatsu S12572-050C}, the signal-to-noise ratio limits the detector lowest operation point for the characterisation. The composition of correlated noise is mainly DiXT and DeXT for the FBK device (\poxeq{7-23} and \pdxeq{6-12} for \dVeq{2\rm{~to~}5}), DiXT and DeXT for the SensL one (\poxeq{7-14} and \pdxeq{3-6} for \dVeq{3\rm{~to~}5}) and only DiXT for the Hamamatsu one (\poxeq{15-34} for \dVeq{1.6\rm{~to~}2.9}).

%% file: gain_measurement.tex
\section{Gain measurement}\label{sec:gain}
The gain of SiPMs is defined as the charge released by a single pixel avalanche. It is independent on the primary source of the avalanche and therefore can be measured with either light or dark noise. The gain is proportional to \dV and to the sum $\rm{C_{d}+C_{q}}$ as defined in the model section~\ref{sec:model}, and can be expressed in units of elementary charge ${\rm G=\Delta V \cdot\left(C_{d}+C_{q}\right)/}e$. Typical values for a detector with $\rm{(50\,\upmu m)^2}$ pixels are $1 - 10\cdot 10^6$ at \dVeq{1 - 7}.
In the following we use G / \dV which has the dimension $\rm{V^{-1}}$ and only depends on detector capacitance. The detector capacitance $\rm{C_{d}+C_{q}}$ is dominated by $\rm{C_{d}}$ which is the pixel capacitance proportional to the pixel active surface and inversely proportional to the avalanche region thickness. It therefore allows to compare the avalanche region thickness of different devices and technologies.
A precise measurement of the gain is important to calculate \fdcr and measure the PDE via the current (see section~\ref{sec:iv} and section~\ref{sec:PDE}).

\subsection{Voltage pulse time-integration}

Measuring the gain with a voltage or charge amplifier requires a calibrated amplifier chain. With our acquisition system presented in section~\ref{exp_methods}, the gain can be determined by a numerical integration of single photon dark or low light injection pulses.
The measured integral $\rm{\int Udt}$ for a single voltage pulse over time leads to the SiPM gain: ${\rm G=1/(R_{load}\cdot G_{Amp}\cdot} e \cdot\int {\rm Udt}$, where the load resistance is typically 50\,$\Omega$ (pre-amplifier input impedance). This method works best with low intensity light pulses. In the off-line analysis we propose to use a fit on single photon pulses (after selection) and proceed with the charge calculation for different \dV. The linear fit of the points allows to extract $\rm{G/\Delta V}$ and evaluate the linearity of the measurement.

\subsection{Current and pulse-frequency measurement}

\begin{figure}[htbp]
\centering
\includegraphics[width=0.55\textwidth]{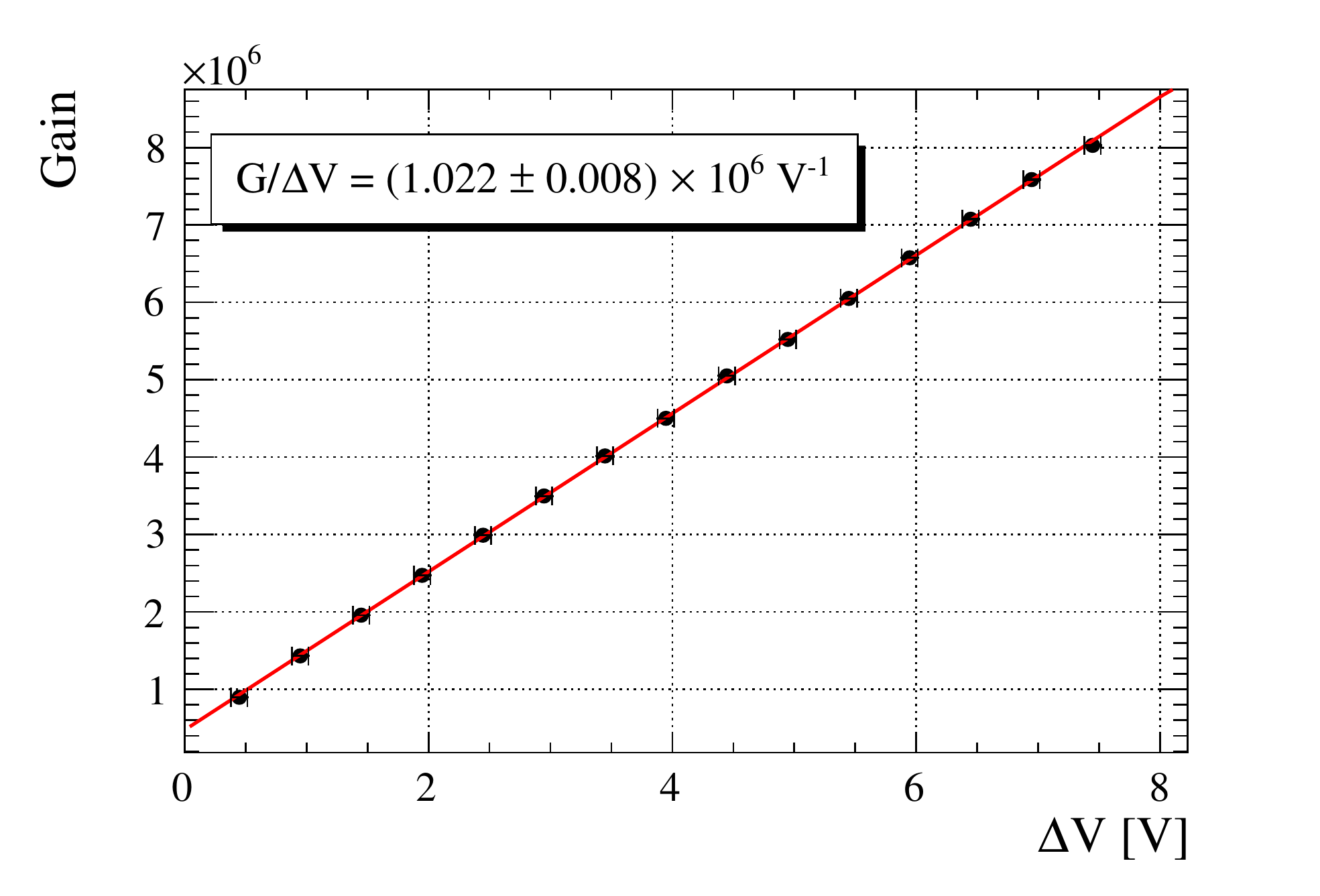}
\caption{Gain as a function of \dV for one channel of an H2017. The slope of the fit corresponds to G/\dV.}
\label{fig:gain_frequency}
\end{figure}

The most accurate gain measurement was obtained by measuring the dark current divided by pulse frequency.
The gain can be calculated as $\rm{I=f_{DCR}\cdot G \cdot} e$.
The assumption of a linear relationship $\rm I \propto G$ holds only in a restricted operation range.
In devices with low \RQ an additional current during the quenching of the avalanche can occur.
Such a current will contribute to a systematic error (non-linearity) in this gain measurement method. Figure~\ref{fig:gain_frequency} shows the calculated gain as a function of \dV for one H2017 channel.
The pulse frequency measurement is corrected for DiXT.
The pulse frequency is measured with a statistical function on the digital oscilloscope.
We employ a statistical method counting peaks in a large interval of $100$\,$\upmu$s with an adjusted threshold.
A correction for the contribution of AP below the threshold was obtained by a numerical integration using AP mean lifetime and recovery time.

%% file: IV_measurement.tex
\section{IV measurements}\label{sec:iv}
We use IV curves measured in reverse direction to obtain the DCR as a function of \dV and to calculate \vbd.
This method provides access to \fdcr as a function of \dV by calculating the ratio of dark current and gain. For devices with high \fdcr, the pulse counting method fails due to the continuous overlap of pulses. In this case, the IV measurement still provides valid results. The IV measurement allows to characterise irradiated detectors with \fdcr above 1\,$\rm{GHz}$/\mm2. Self heating due to the high bias voltage current for devices with high \fdcr can limit the voltage scan range. The measurement of \vbd is not affected by self heating because only the region with small currents close to \vbd needs to be scanned.
The IV curve taken in forward direction allows to measure the quench resistor.

\subsection{Dark count rate}
With the gain obtained in section~\ref{sec:gain}, \fdcr can be calculated with the relation: $\rm{I} = \rm{f_{DCR}}\cdot\rm{G}\cdot e\cdot (1 + \rm{p_{DiXT}})$ where I is the current obtained from the IV measurement.
The DCR obtained in this way accounts for all random dark pulses and for the delayed correlated pulses. The correction for DiXT is applied.

\begin{figure}[htbp]
\centering
\subfloat[The IV curve in forward direction is fitted with a linear function to obtain the values of \RQ. \label{fig:RqfitandRq}]{\includegraphics[width=0.47\textwidth]{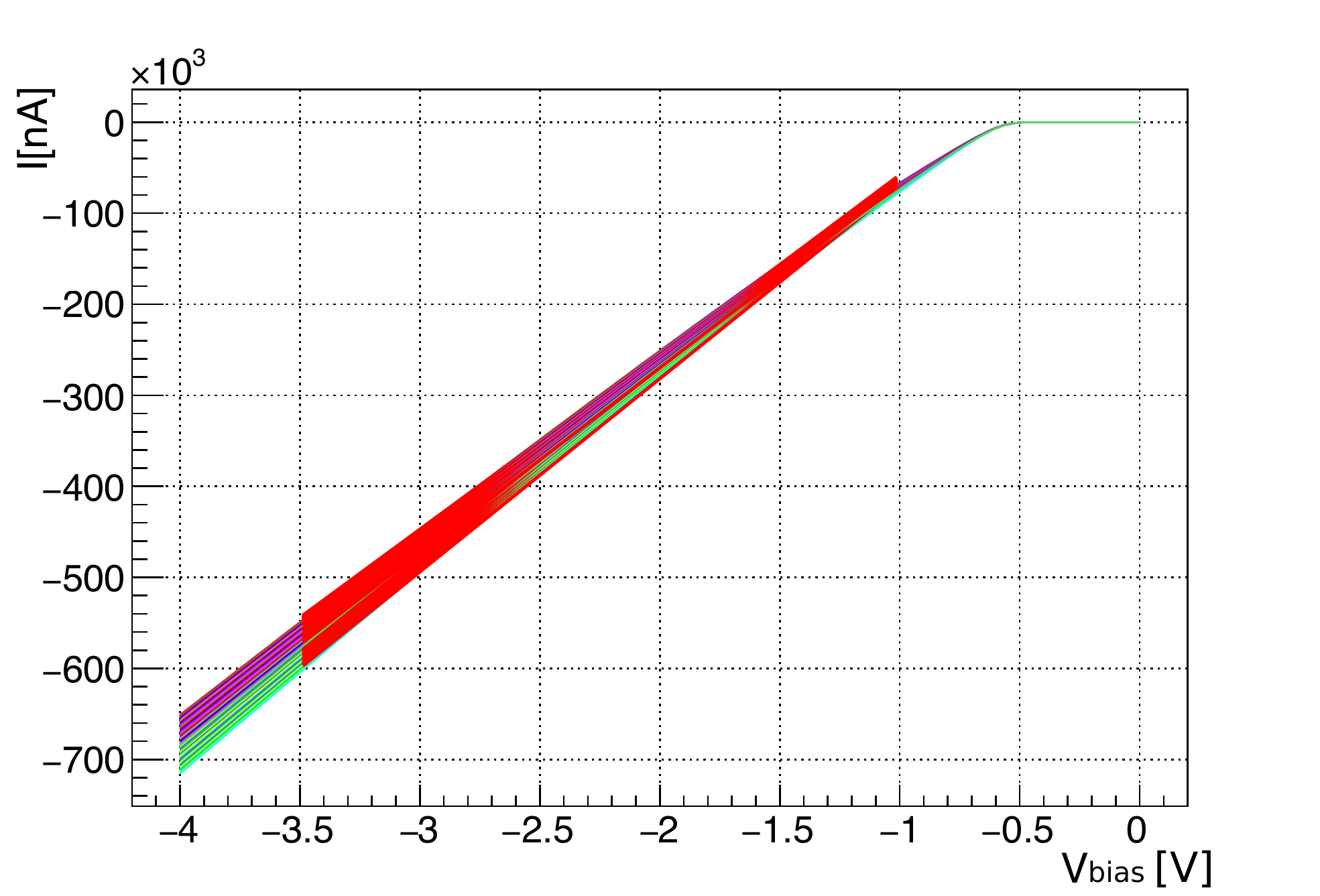}}\qquad
\subfloat[The IV curves in reverse direction near \vbd for a selection of 4 channels is shown. ]{\includegraphics[width=0.47\textwidth]{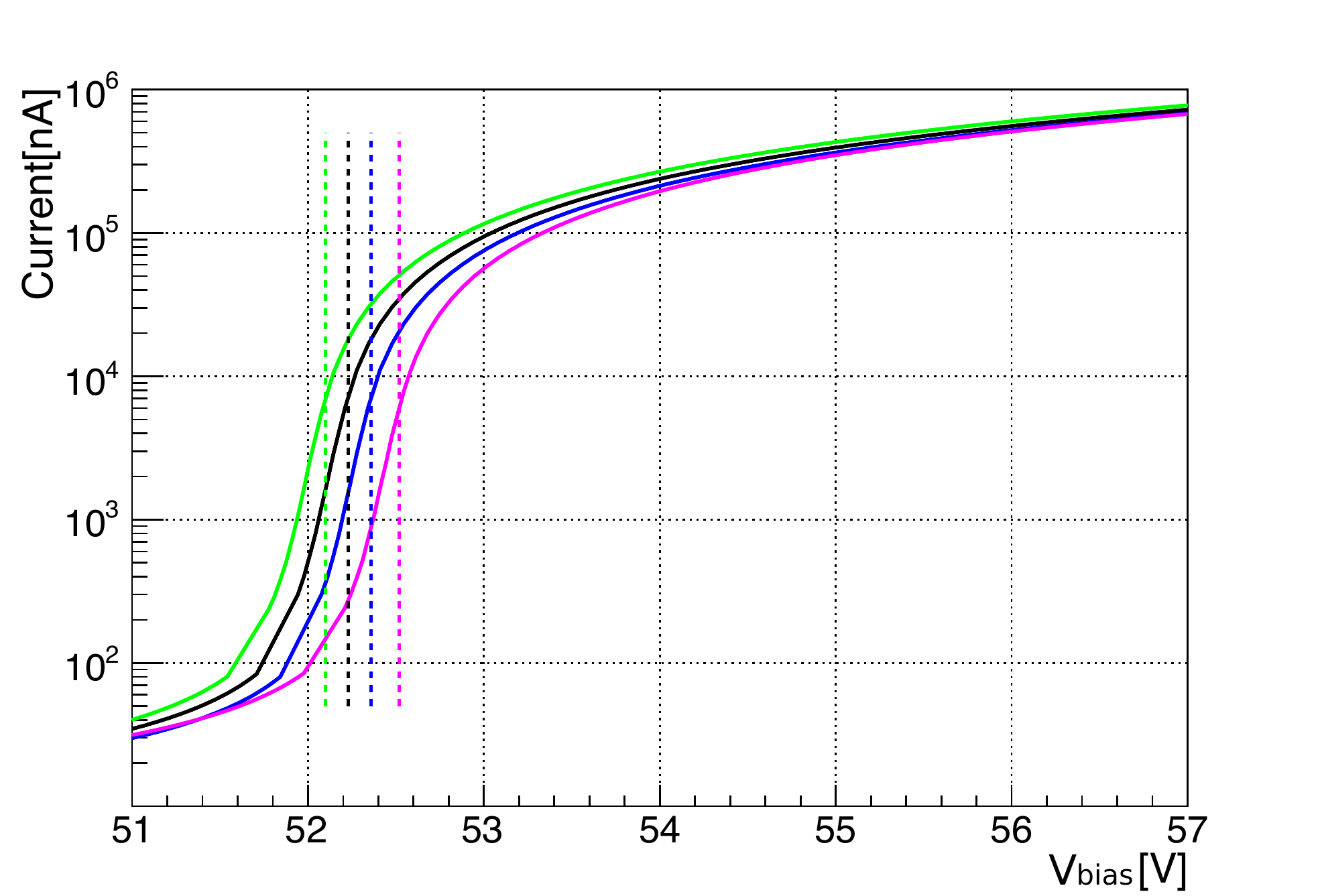}}\qquad
\subfloat[The value for \RQ of the 128 channel array is shown. ]{\includegraphics[width=0.47\textwidth]{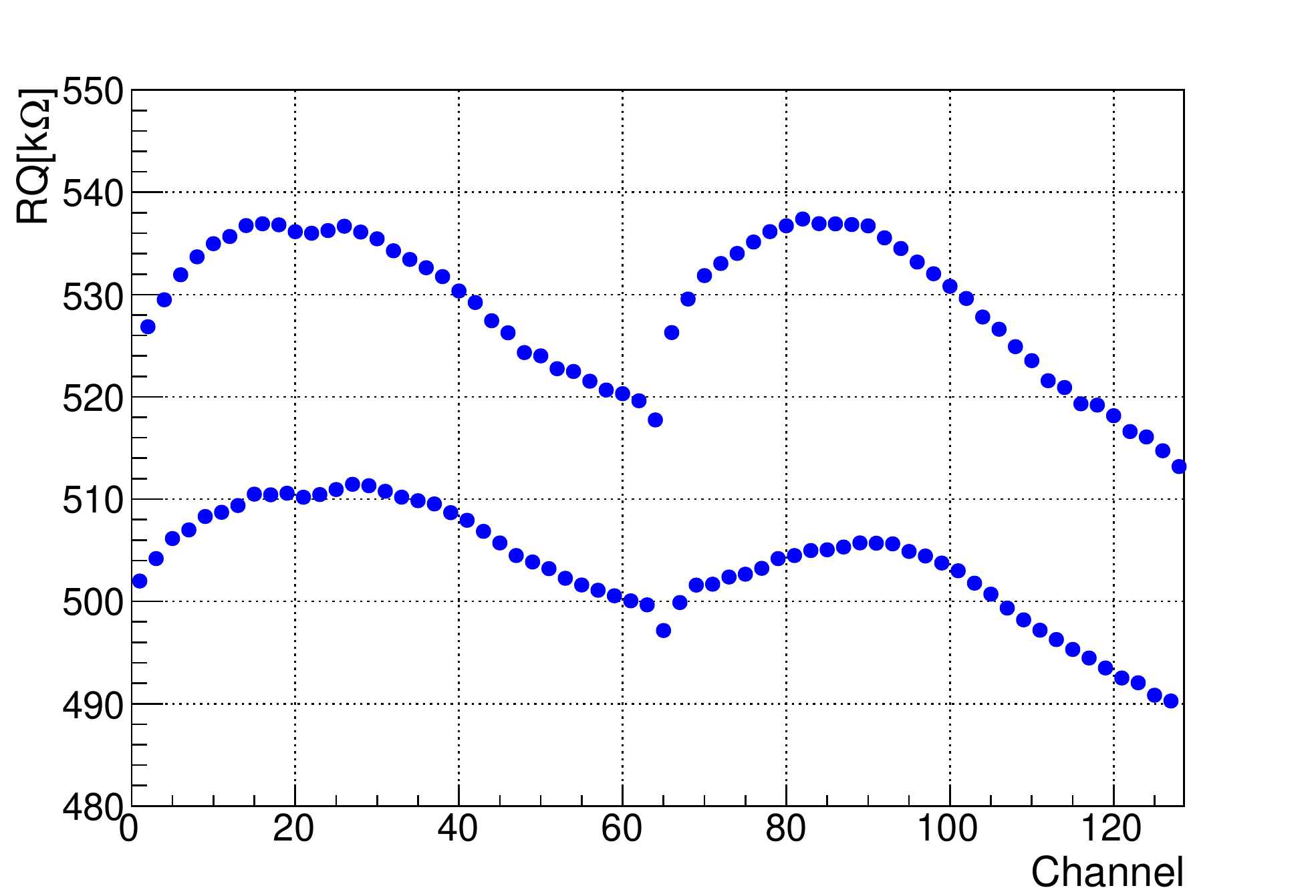}} \qquad
\subfloat[ The linear fit and the extrapolation to zero allows to calculate \vbdI. \label{fig:IVandDeriv}]{\includegraphics[width=0.47\textwidth]{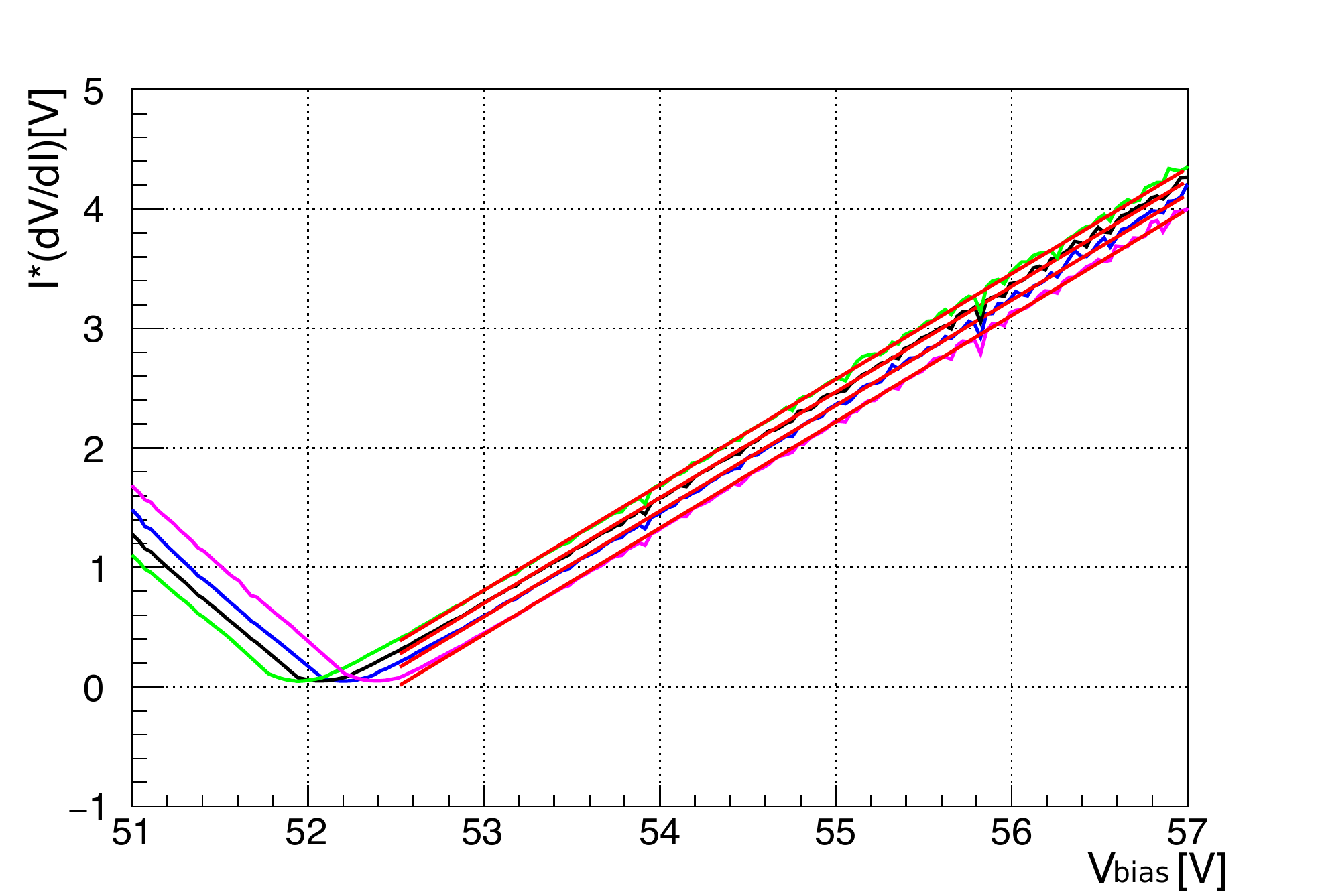}}
\caption{The IV curves in forward and reverse direction allow to measure \RQ and \vbdI. The measured values for \vbdI are indicated with the vertical lines in (b). }
\end{figure}

\subsection{Breakdown voltage}\label{sec:IV-VBD}
From the reverse bias region where a rapid current increase is seen, the \vbd can be extracted. Reference~\cite{IV_Garuti} presents a method to determine \vbd from the IV curve. The assumption of an exponential current increase leads to $\rm{I\left(V_{bias}\right)=\alpha \cdot \left(V_{bias}-V_{BD}\right)^\varepsilon}$ where I is the measured current, and $\alpha$ and $\varepsilon$ are constants that determine the shape of the IV curve. The linear fit of:
\begin{equation}
\label{inv_iv}
\rm{\left[\frac{dln(I)}{dV_{bias}}\right]^{-1} = \frac{\left(V_{bias}-V_{BD}^I\right)}{\varepsilon}}
\end{equation}
is used to determine \vbdI as illustrated in Fig.~\ref{fig:IVandDeriv}. The subscript "I" indicates that the quantity \vbdI is different from \vbdG  obtained by the gain measurement in section~\ref{sec:gain}.
The difference observed $\rm{V_{BD}^I-V_{BD}^G}$ is $320$\,mV and is found to be constant for the H2017 detectors in the temperature range of 25\degC to -40\degC. This difference was reported by~\cite{TurnOn_TurnOff} and found to depend on the pixel size. The observed variations between channels on a large sample is smaller than 10\,mV. To improve the signal-to-noise ratio for \vbd measurement, a constant light source can be used without affecting the \vbd value.

\subsection{Quench resistor}
The IV characteristics in the region [0,-1]\,V show the typical shape of a forward biased pn-junction. For higher forward voltages the current is limited by \RQ and shows an almost linear characteristic, as shown in Fig.~\ref{fig:RqfitandRq}. A linear fit is applied in the region from -1\,V to -3.5\,V and the slope extracted leads to: $\rm{R_Q=N_{pixel} \cdot dV/dI}$, where $\rm{N_{pixel}}$ is the total number of pixels and $\rm dV/dI$ the fitted slope. Note that the result depends on the fit region due to the non linear characteristics of the pn-junction in series.

%% file: pde_measurement.tex
\section{PDE measurement}\label{sec:PDE}
The measurement of the absolute PDE requires a precise determination of the gain and the correction for the correlated noise. For devices with low \fdcr, the comparison of the PDE among different devices (relative PDE) can be measured based on zero photon probability in a low light amplitude spectrum\,\footnote{~This method is based on the Poisson probability  $\rm{P(0,\mu)=\exp^{-\upmu}}$ where $\upmu$ is the mean value of number of photons detected. Since only the component without photons needs to be measured, the gain and noise corrections are not required. Keeping the light intensity unchanged, the comparison of different devices is possible.}. In the following, two independent absolute measurements are described and the corrections discussed.

\subsection{Light beam intensity}
As described in section~\ref{exp_methods}, the PDE measurement setup provides a uniform and calibrated low intensity light beam. The light intensity is kept sufficiently low such that saturation or recovery of pixels can be neglected. For a device with recovery time of $\rm \tau_{rec}=84$\,ns, the rate of detected photons $\rm{R=1/(3\cdot\tau_{rec})\approx 4\,MHz}$ should not be exceeded. The pulse frequency (counting method) described below, is a threshold based counting method. It is well suited for devices with low number of pixels because it is based on single pulse counting. For large devices, the measurement method based on the current is more suitable.

\subsection{PDE measurement based on the current method}
The measured photo-current of an SiPM in a light beam is in first approximation ${\rm I=R\cdot G\cdot} e$. In reality the recorded current $\rm{I_{recorded}}$ is the sum of the photo-current $\rm{I_{light}^*}$\,\footnote{~In the following the asterisk indicates  quantities that also include correlated noise.} and the dark current, $\rm{I_{dark}^*}$.
Both contributions can be corrected for any type of correlated noise by a correction factor that we call $\left(1-\rm{r_{curr}}\right)$.
DiXT and DeXT contribute to the current with pulses of $1$\,PE and AP with a fraction of $1$\,PE pulses.
To correct for the variable amplitude of AP, we define $\rm{w_{AP}}$ as the mean fraction of a $1$\,PE pulse such that $\rm{w_{AP}\cdot p_{AP}}$ is equal to the total charge released from AP.

The correction for the correlated noise can be made accurate for high \dV where correlated noise probabilities are above $10\%$, only if the correction for higher order correlated noise are introduced (e.g. AP of DiXT and AP of DeXT).
In the waveform analysis the higher order corrections are calculated for this purpose~(see Fig.~\ref{primary_secondary_corr_noise}). The average charge $\rm{w_{h.o.}}$ in such pulses and the probability $\rm{p^{curr}_{h.o.}}$ of their occurrence are made available\,\footnote{~Note that the DiXT probability is accounting for single and higher order DiXT. The mean amplitude can be used instead of a 1\,PE if necessary.}.
The relation between the current measured on the SiPM and the different noise contributions are expressed by the following equation:
\begin{equation}\label{eq:I_corr}
\rm{I^* = I\cdot \left[1 + \left(1\cdot p_{DeXT}\right) + \left(1\cdot p_{DiXT}\right) + \left(w_{AP}\cdot p_{AP}\right) + \left(w_{h.o.}\cdot p^{curr}_{h.o.}\right)\right]}.
\end{equation}
Eq.~\ref{eq:I_corr} is valid for a current produced either by light or by dark noise. We can invert this relation for $\rm I^*$ and $\rm I$ and find the current produced by light via one single correction factor $\rm{r_{curr}}$:
\begin{equation}
\rm{I_{light} = I_{light}^*\cdot \left(1-r_{curr}\right)}.
\end{equation}
The current produced by light is also equal to the total recorded current corrected by the dark current leading to:
\begin{equation} \label{eq:I_light}
\rm{I_{light}  = \left(I_{recorded} - I_{dark}^*\right)\cdot \left(1-r_{curr}\right)},
\end{equation}
where $\rm{I_{recorded}}$ and $\rm{I_{dark}^*}$ are measured respectively with injected light and in the dark. The correction factor, $\rm r_{curr}$, is calculated from the waveform analysis results. The most important aspects are discussed below.

\paragraph{AP}
To gain robustness for an automated measurement, the threshold level for pulse detection is set to 0.6\,PE. To fully account for the current produced by AP, the following weight is applied: $\rm{w_{AP}^{0.6}=7.2}$ resulting from a numerical integration  described in paragraph~\ref{ap_lifetimes}.
\paragraph{Multiple direct cross-talk}
A threshold of 1.17\,PE is used to detect DiXT by its amplitude. The detection of multiple DiXT is not implemented. This contribution is neglected in the current version. The systematic error is estimated to be 1\% for a \pox=10\%.

\paragraph{DiXT on delayed pulses} The amplitude of delayed 1\,PE pulses are baseline shifted with the slow component of the primary pulse. DiXT produced by these pulses cannot easily be discriminated by their amplitude. As a consequence, the effect of DiXT on delayed pulses is currently not taken into account for the corrections. The expected error is of the order of 0.5\% for a total correlated noise probability of 10\%.

With the above considerations, $\rm{r_{curr}}$ can be extracted from the waveform analysis using the following relation:
\begin{equation}\label{eq:r_curr}
\rm{r_{curr} = \frac{N_{delayed\,pulses}}{N_{ev}+N_{delayed\,pulses}} + \left(w_{AP}^{0.6}-1\right)\cdot p_{AP}^{0.6} + p_{DiXT}},
\end{equation}
where $\rm{N_{delayed\,pulses}}$ is the number of delayed pulses exceeding the threshold and $\rm{N_{ev}}$ is the total number of waveforms recorded.

Given the photo-current corrected for noise as in Eq.~\ref{eq:I_light}, the SiPM gain and the ratio of the active surfaces the PDE can finally be computed:
\begin{equation}\label{PDE_curr}
\rm PDE = QE_{PD}\cdot\frac{I_{light}}{G\cdot I_{PD}}\cdot\frac{A_{PD}}{A_{SiPM}},
\end{equation}
where $\rm A_{PD}$ and $\rm A_{SiPM}$ are respectively the surface of the photodiode and the tested SiPM.

\paragraph{Statistical and systematic uncertainties}
We have identified two dominating systematic uncertainties related to the setup and three to the correction and the gain computation. The photo-diode calibration has a 1\% precision leading to a 1\% uncertainty. This value is given for the peak wavelength. In addition, the accuracy of the positioning of the photo-diode and the SiPM in the light beam introduces the second major uncertainty. The accuracy of the positioning along the beam axis is crucial since the intensity depends quadratically on the distance. Assuming $\rm d = 200\pm 2$\,mm, a 2\% uncertainty is obtained. We have evaluated also the homogeneity in the plane perpendicular to the light beam and found that for a position accuracy of $\pm 0.5$\,mm the total systematic uncertainty is smaller than 1$\%$. With a multichannel array we repeated the measurement for different channels displaced by 4\,mm and confirmed the homogeneity measurement.

On the correlated noise computation side, the corrections suffer from growing systematic errors as correlated noise increases. We estimate that for a total correlated noise below 10\% the correction has a systematic uncertainty that influences the PDE by less than 1\%. The uncertainty is expected to increase with correlated noise.
Finally, a dominant contribution to the uncertainty of the current based PDE measurement is linked to the gain measurement. With the method described in section~\ref{sec:gain}, an accuracy of 1\% can been achieved. In addition, to achieve good gain stability to the level of 1\%, the
 temperature needs to be stabilised to 0.5\,K over the full measurement period\,\footnote{~For H2017 detectors operated at $\rm{\Delta V=3.5\,V}$, with $\rm{K_{V_{BD}}=56\,mV/^\circ C}$, a change in temperature of $0.5$\degC results in a gain change of $0.8\%$.}.
 
In the wavelength regions with high photon sensitivity, total systematic uncertainty is dominant over the statistical one, while in low sensitivity regions and low \dV, the statistical dominates. As a summary, the current based PDE method has a total estimated uncertainty of $\sim$6\%, where one third is caused by the gain measurement.

\subsection{PDE measurement based on the pulse frequency method}
To overcome the large error introduced by the gain measurement we propose to perform the PDE measurement based on pulse counting.
This method depends on a low pulse frequency such that random overlap of pulses is rare and therefore cannot significantly affect the precision. For this method, the frequency of pulses $\rm{f_{recorded}}$ at a threshold of $0.6$\,PE is measured. The frequency obtained in this way, is the sum of the light induced pulses, $\rm f_{light}^*$, and dark pulses, $\rm f_{dark}^*$. The asterisk has been placed to indicate that these values also contain a fraction of correlated noise induced pulses:
\begin{equation}\label{f_tot}
\rm{f^* = f\cdot \left(1 + p_{AP}^{0.6} + p_{DeXT} + p^{freq}_{h.o.}\right)},
\end{equation}
where $\rm f$ is the frequency of detected primary pulses (light induced or dark count) and $\rm{p^{freq}_{h.o.}}$ is the probability for higher order delayed correlated noise to occur. The frequency of primary photons $\rm{f_{light}}$ is therefore computed as:
\begin{equation}
\rm{f_{light} = f_{light}^*\cdot \left(1-r_{freq}\right) = \left(f_{recorded} - f_{dark}^*\right)\cdot \left(1-r_{freq}\right)}.
\end{equation}
The parameter $\rm{r_{freq}}$ is the fraction of pulses due to delayed correlated noise. The DiXT pulses are only counted as one pulse and do not require any correction. Precise values  for $\rm{r_{freq}}$ are obtained from the pulse-shape analysis applied on dark pulses. It is computed as:
\begin{equation}
\rm{r_{freq} = \frac{N_{delayed\,pulses}}{N_{ev}+N_{delayed\,pulses}}}.
\end{equation}
The PDE of the SiPM can be calculated as:
\begin{equation}\label{PDE_freq}
\rm PDE = QE_{PD}\cdot\frac{f_{light}}{I_{PD}/e}\cdot\frac{A_{PD}}{A_{SiPM}},
\end{equation}
which uses the current of the photo-diode $\rm I_{PD}$ for the calculation.

\paragraph{Statistical and systematic uncertainties}
The sources of uncertainties for the setup are identical as for the current-based method. Nonetheless, the systematic error benefits from two large improvements. First, the correction for correlated noise is computed simply by counting delayed pulses and the DiXT probability is not used. Second, the gain is not used in the PDE calculation which eliminates its error as well as the dependence on temperature. In summary, the total systematic and statistical uncertainties are reduced to $\sim$3.5\%.

\subsection{Results}
\begin{figure}[htbp]
\centering%
\subfloat[Current based method \label{fig:curwav}]{\includegraphics[width=0.5\textwidth]{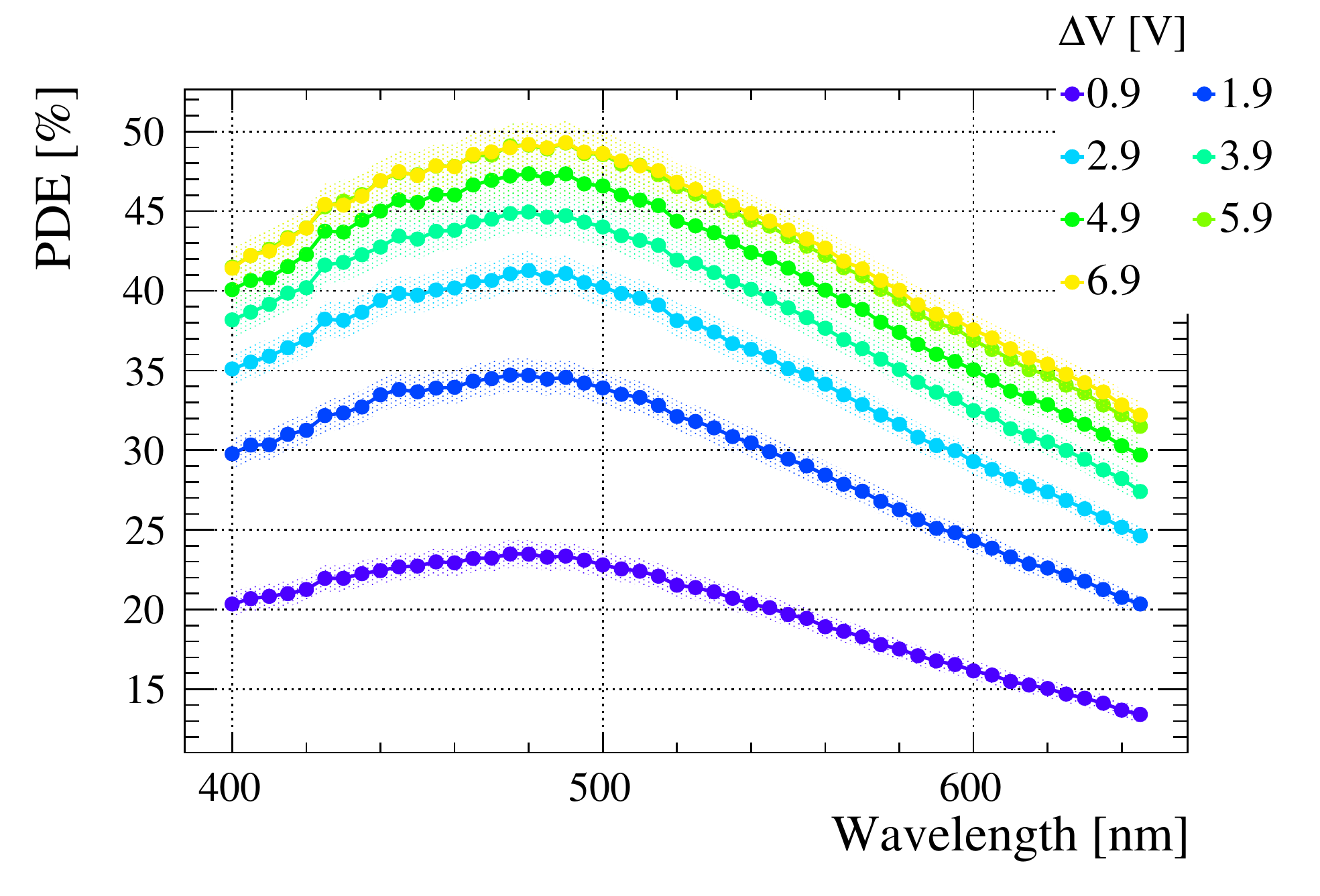}}\hfill
\subfloat[Current based method \label{fig:curbias}]{\includegraphics[width=0.5\textwidth]{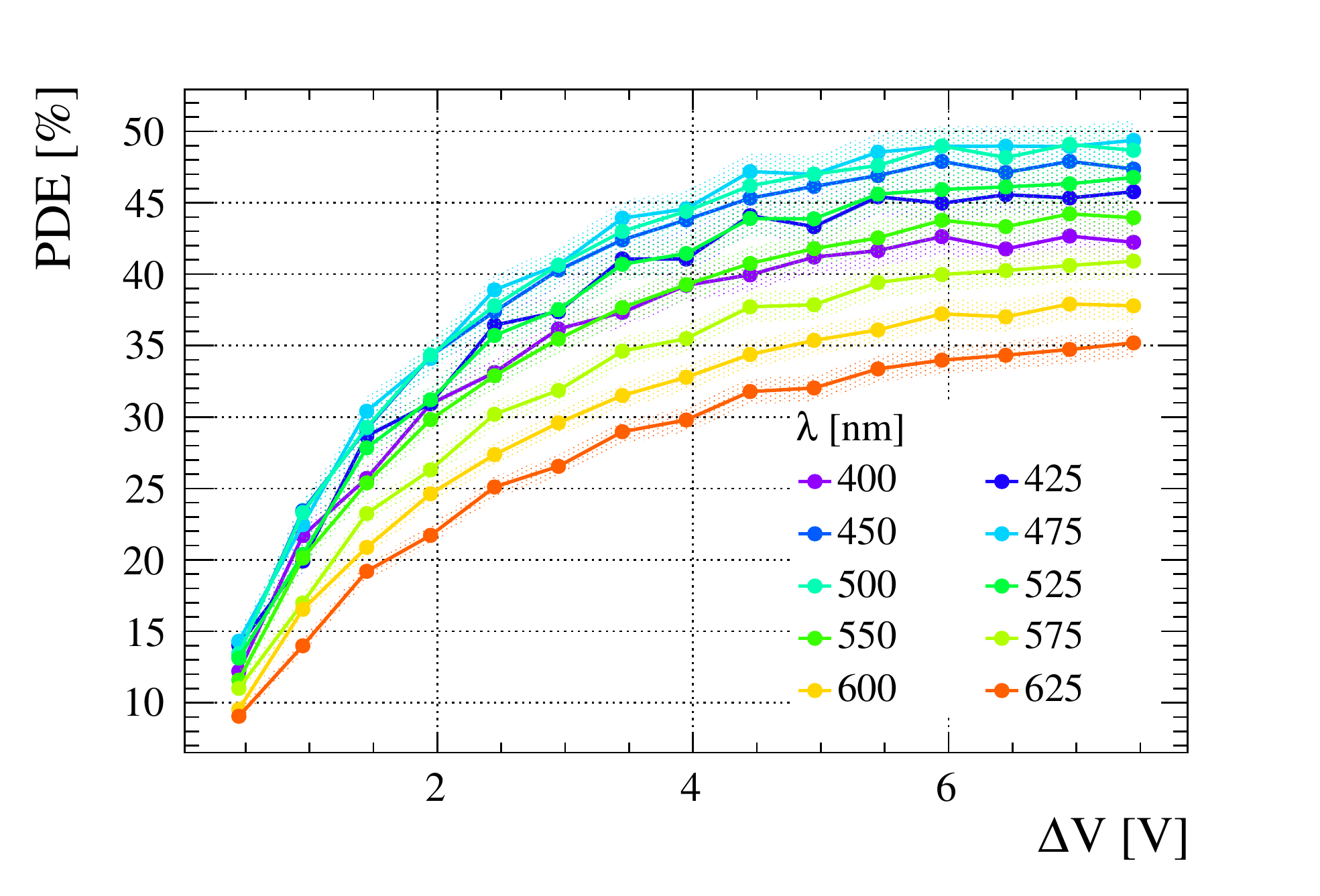}}\hfill
\subfloat[Frequency based method
\label{fig:freqwav}]{\includegraphics[width=0.5\textwidth]{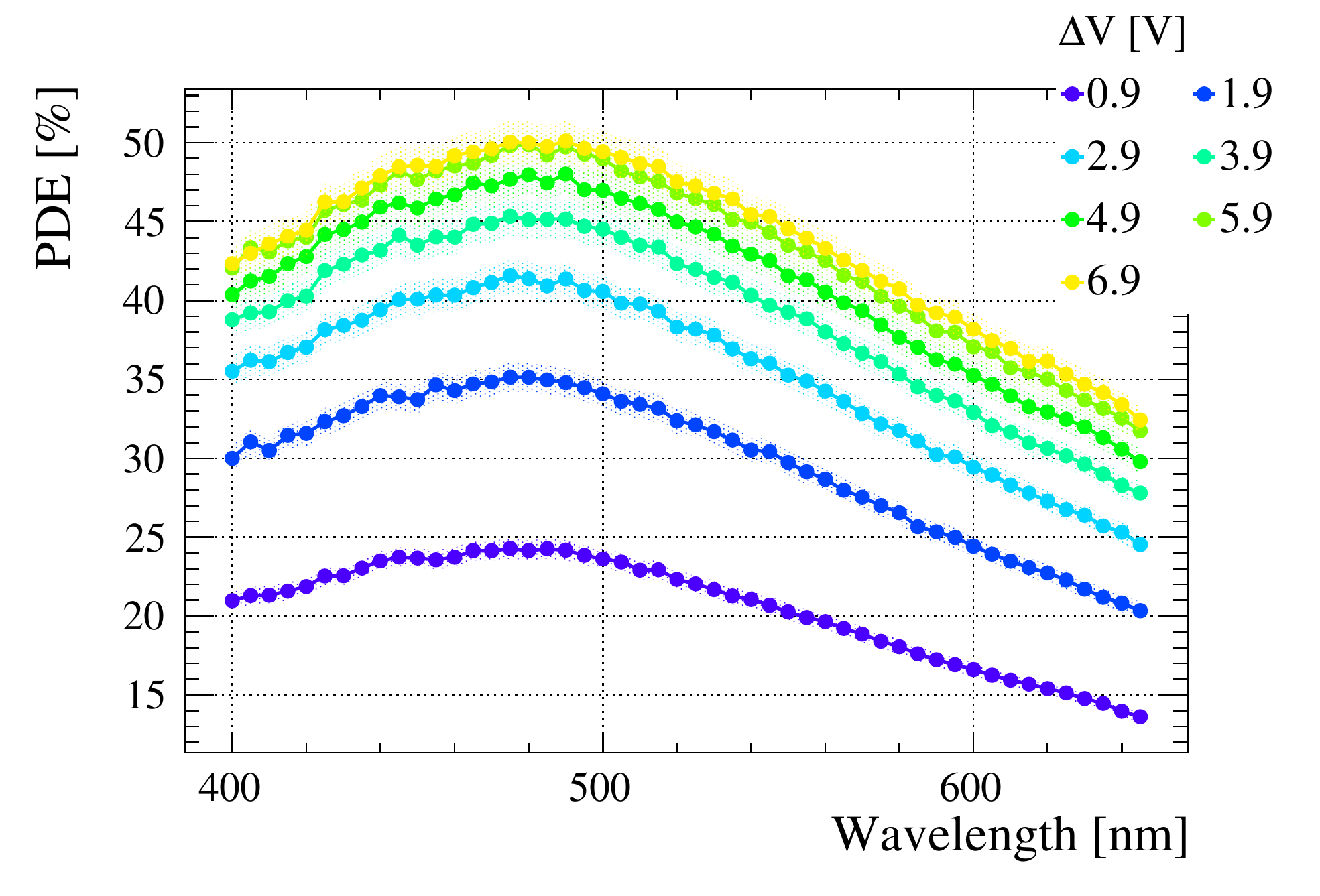}}\hfill
\subfloat[Frequency based method \label{fig:freqbias}]{\includegraphics[width=0.5\textwidth]{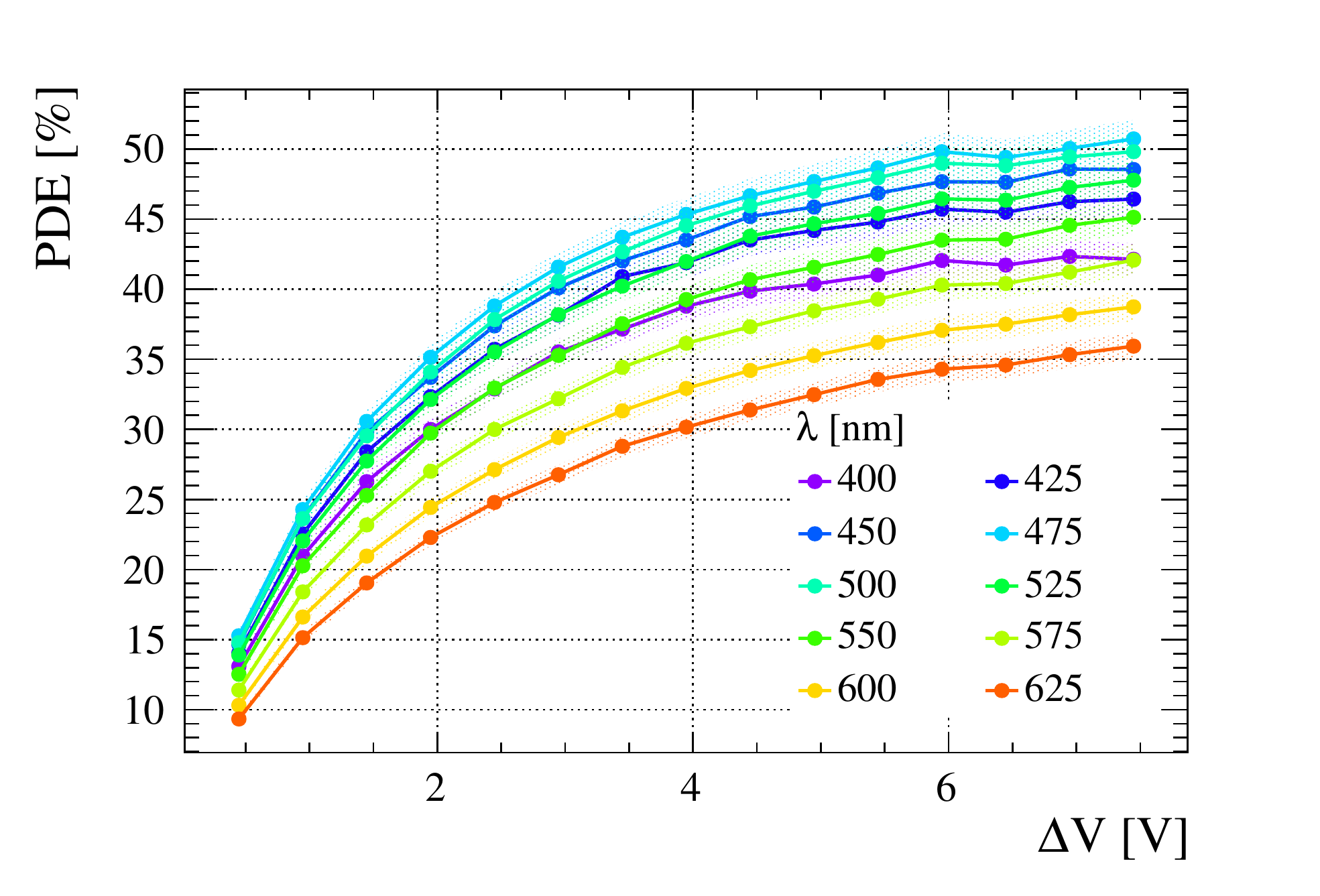}}
\caption{PDE measurement for an H2017 detector, in (a) and (c) different curves correspond to different \dV and in (b) and (d) to different wavelengths.}
\label{fig:PDE_results}
\end{figure}

The PDE as a function of wavelength and \dV obtained with the two methods for an H2017 detector are shown in Fig.~\ref{fig:curwav}, \ref{fig:curbias}, \ref{fig:freqwav} and \ref{fig:freqbias}. The absolute values have been cross checked with Hamamatsu HPK and presents less than 2\% difference at the nominal operation voltage.

%% file: conclusion.tex
\section{Conclusion}
In the context of the LHCb SciFi tracker and the customisation of multichannel SiPM arrays, methods to evaluate SiPMs were developed.  The  characterisation is based on time dependent waveforms recorded by an oscilloscope and IV scans. A framework based on the ROOT library has been employed to analyse and handle the classification of different types of correlated noise based on the pulse-shapes.
The method has been successfully applied to devices with pixel size as small as $\rm{(25\,\upmu m)^2}$.

The measured parameters are: breakdown voltage, direct and delayed cross-talk probabilities, after-pulse probability, absolute photon detection efficiency, after-pulse and delayed cross-talk mean lifetimes, recovery time and the long pulse component time constant, quench resistor and dark count rate.

The methods have been illustrated using measurements from the the final LHCb SciFi device from Hamamatsu (H2017 multichannel array). Results with devices from other manufactures were performed and summarised (KETEK $1.625\times 0.250$\,\mm2 (multichannel array), SensL $3\times 3$\,\mm2 and FBK $1\times 1$\,\mm2).

%% file: main.bbl
\begin{thebibliography}{1}

\bibitem{TDR}
LHCb Collaboration.
\newblock {LHCb Tracker Upgrade Technical Design Report}.
\newblock 2014.

\bibitem{root_cern}
Rene Brun and Fons Rademakers.
\newblock Root - an object oriented data analysis framework.
\newblock {\em Proceedings AIHENP'96 Workshop, Lausanne, Sep. 1996, Nucl. Inst.
  $\&$ Meth. in Phys. Res. A 389}, 1997.

\bibitem{Calouzol_FBK_cryo}
G.~Collazuol, M.G. Bisogni, S.~Marcatili, C.~Piemonte, and A.~Del Guerra.
\newblock Studies of silicon photomultipliers at cryogenic temperatures.
\newblock {\em Nuclear Instruments and Methods in Physics Research Section A:
  Accelerators, Spectrometers, Detectors and Associated Equipment}, 628(1):389
  -- 392, 2011.
\newblock VCI 2010.

\bibitem{VATA64}
M.G.~Bagliesi et~al.
\newblock A custom fron-end asic for the readout and tinming of 64 sipm
  photosensors.
\newblock {\em Nuclear Physics B Proc. Suppl.) 215 (2011) 344-348}, 2011.

\bibitem{Corsi_Asic_development_for_SiPM_readout}
Francesco Corsi, Maurizio Foresta, Cristoforo Marzocca, Gianvito Matarrese, and
  Alberto~Del Guerra.
\newblock Asic development for sipm readout.
\newblock {\em Journal of Instrumentation}, 4(03):P03004, 2009.

\bibitem{IV_Garuti}
E.Garutti et~al.
\newblock Characterization and x-ray damage of silicon photomultipliers.
\newblock {\em Proceedings of Science, proceedings to talk at Technology and
  Instrumentation in Particle Physics 2014}, 2-6 June 2014.

\bibitem{TurnOn_TurnOff}
G.~Zhang et~al.
\newblock Turn-on and turn-off voltages of an avalanche p-n junction.
\newblock {\em J. Semicond.}, vol. 33(no. 9, p. 094003), Sep. 2012.

\end{thebibliography}
